\documentstyle[emulateapj,epsf]{article}
\begin{document}

\newcommand{\beq}{\begin{equation}}
\newcommand{\eeq}{\end{equation}}
\newcommand{\beqn}{\begin{eqnarray}}
\newcommand{\eeqn}{\end{eqnarray}}
\newcommand{\pa}{\partial}
\newcommand{\vp}{\varphi}
\newcommand{\varep}{\varepsilon}
\newcommand{\ep}{\epsilon}

\lefthead{SHIBATA, BAUMGARTE \& SHAPIRO}
\righthead{BAR-MODE INSTABILITY IN DIFFERENTIALLY ROTATING NEUTRON STARS}

\title{THE BAR-MODE INSTABILITY IN DIFFERENTIALLY ROTATING NEUTRON STARS:
	SIMULATIONS IN FULL GENERAL RELATIVITY}

\author{Masaru Shibata\altaffilmark{1}, Thomas W. Baumgarte, and 
Stuart L. Shapiro\altaffilmark{2}}

\affil{Department of Physics, University of Illinois at
        Urbana-Champaign, Urbana, Il~61801}

\altaffiltext{1}{Department of Earth and Space Science, Graduate School of
	Science, Osaka University, Toyonaka, Osaka 560-0043, Japan}

\altaffiltext{2}{Department of Astronomy and NCSA, 
	University of Illinois at Urbana-Champaign, Urbana, IL 61801}

\begin{abstract}
We study the dynamical stability against bar-mode deformation of
rapidly spinning neutron stars with differential rotation.  We perform
fully relativistic 3D simulations of compact stars with $M/R \geq 0.1$, 
where $M$ is the total gravitational mass and $R$ the equatorial
circumferential radius. We adopt an adiabatic equation of state with
adiabatic index $\Gamma=2$.  As in Newtonian theory, we find that
stars above a critical value of $\beta \equiv T/W$ (where $T$ is the
rotational kinetic energy and $W$ the gravitational binding energy) are
dynamically unstable to bar formation. For our adopted choices of
stellar compaction and rotation profile, the critical value of $\beta
= \beta_{dGR}$ is $\sim 0.24-0.25$, only slightly smaller than the
well-known Newtonian value $\sim 0.27$ for incompressible Maclaurin
spheroids.  The critical value depends only very weakly on the degree
of differential rotation for the moderate range we surveyed.  All
unstable stars form bars on a dynamical timescale.  Models with
sufficiently large $\beta$ subsequently form spiral arms and eject
mass, driving the remnant to a dynamically stable state.  Models with
moderately large $\beta \gtrsim \beta_{dGR}$ do not develop spiral
arms or eject mass but adjust to form dynamically stable
ellipsoidal-like configurations.  If the bar-mode instability is
triggered in supernovae collapse or binary neutron star mergers, it
could be a strong and observable source of gravitational waves.  We
determine characteristic wave amplitudes and frequencies.
\end{abstract}

\section{INTRODUCTION}

Neutron stars in nature are rotating and subject to nonaxisymmetric
rotational instabilities.  An exact treatment of these instabilities
exists only for incompressible equilibrium fluids in Newtonian
gravity (see, e.g., Chandrasekhar 1969; Tassoul 1978; 
Shapiro \& Teukolsky 1983). For these configurations, global
rotational instabilities arise from nonradial toroidal modes
$e^{im\varphi}$ ($m=\pm 1,\pm 2, \dots$) when $\beta\equiv T/W$ exceeds a
certain critical value. Here $\varphi$ is the azimuthal coordinate and
$T$ and $W$ are the rotational kinetic and gravitational potential binding
energies.  In the following we will focus on the $m=\pm 2$ bar mode,
since it is the fastest growing mode when the rotation is sufficiently
rapid.

There exist two different mechanisms and corresponding timescales for
bar mode instabilities.  Uniformly rotating, incompressible stars in
Newtonian theory are {\em secularly} unstable to bar mode formation
when $\beta \geq \beta_s \simeq 0.14$.  However, this instability can
only grow in the presence of some dissipative mechanism, like
viscosity or gravitational radiation, and the growth time is
determined by the dissipative timescale, which is usually much longer
than the dynamical timescale of the system.  By contrast, a {\em
dynamical} instability to bar mode formation sets in when $\beta \geq
\beta_d \simeq 0.27$.  This instability is independent of any
dissipative mechanisms, and the growth time is determined by the
hydrodynamical timescale of the system.

The secular instability in compressible stars, both uniformly and
differentially rotating, has been analyzed numerically within linear
perturbation theory, by means of a variational principle and trial
functions, by solving the eigenvalue problem, or by other approximate
means.  This technique has been applied not only in Newtonian theory
(Lynden-Bell \& Ostriker 1967; Ostriker \& Bodenheimer 1973; Ipser and
Lindblom 1989; Friedman and Schutz 1977) but also in post-Newtonian
theory (Cutler and Lindblom 1992; Shapiro and Zane 1998 for
incompressible stars) and full general relativity (Yoshida and
Eriguchi 1995; Bonazzola, Frieben and Gourgoulhon 1996; Stergioulas
and Friedman 1998). For relativistic stars, the critical value of
$\beta_s$ depends on the compaction $M/R$ of the star (where $M$ is
the gravitational mass and $R$ the circumferential radius at the equator),
on the rotation law and on the dissipative mechanism.  The
gravitational-radiation driven instability occurs for smaller rotation
rates, i.e. for values $\beta_s < 0.14$, in general relativity.  For
extremely compact stars (Stergioulas and Friedman 1998) 
or strongly differentially rotating stars (Imamura et al. 1995), 
the critical value can be as small as $\beta_s < 0.1$. 
By contrast,
viscosity drives the instability to higher rotation rates $\beta_s >
0.14$ as the configurations become more compact (Bonazzola, Frieben
and Gourgoulhon 1996; Shapiro and Zane 1998).

Determining the onset of the dynamical bar-mode instability, as well
as the subsequent evolution of an unstable star, requires a numerical
simulation of the fully nonlinear hydrodynamical equations.
Simulations performed in Newtonian theory (e.g.~Tohline, Durisen \&
McCollough 1985; Durisen, Gingold \& Tohline 1986; Williams \& Tohline
1988; Houser, Centrella \& Smith 1994; Smith, Houser \& Centrella
1995; Houser \& Centrella 1996; Pickett, Durisen \& Davis 1996; New,
Centrella \& Tohline 1999) have shown that $\beta_d$ depends only very
weakly on the stiffness of the equation of state.  Once a bar has
developed, the formation of spiral arms plays an important role in
redistributing the angular momentum and forming a core-halo structure.
Recently, it has been shown that, similar to the onset of secular
instability, $\beta_d$ can be smaller for stars with a higher degree
of differential rotation (Tohline \& Hachisu 1990; Pickett, Durisen \&
Davis 1996)

To date, the dynamical bar-mode instability has been analyzed only in
Newtonian theory, hence almost nothing is known about the role of
relativistic gravitation.  The reason is that until quite recently a
stable numerical code capable of performing reliable hydrodynamic
simulations in three dimensions plus time in full general relativity
has not existed.  Some recent developments, however, have advanced the
field significantly.  New formulations of the Einstein equation based
on modifications of the standard $3+1$ ADM system of equations
(Arnowitt, Deser \& Misner 1962) have resulted in codes which have
proven to be remarkably stable over many dynamical timescales (e.g.,
Shibata \& Nakamura 1995; Baumgarte \& Shapiro 1999; Oohara \&
Nakamura 1999).  In addition, gauge conditions which warrant long-time
stable evolution for rotating and self-gravitating systems and are
manageable computationally have been developed (e.g., Shibata 1999b).
In this paper we adopt the relativistic hydrodynamic implementation of
Shibata (1999a) to study the onset and growth of the dynamical
bar-mode instability in relativistic stars.  Although this study is
carried out only for a simple equation of state and rotational law, it
demonstrates how, as numerical relativity in full $3+1$ matures, it is
becoming more useful as a tool to solve long-standing problems in
relativistic astrophysics characterized by strong gravitational fields
and little or no spatial symmetry.

There are numerous evolutionary paths which may lead to the formation
of rapidly rotating neutron stars with $\beta \sim 0.3$. The parameter
$\beta$ increases approximately as $R^{-1}$ during stellar collapse.
During supernova collapse, the core contracts from $\sim 1000$ km to
$\sim 10$ km, and hence $\beta$ increases by about two orders of
magnitude.  Thus, even moderately rapidly rotating progenitor stars
may yield rapidly rotating neutron stars which may reach the onset of
dynamical instability (Bonazzola \& Marck 1993; Rampp, M\"uller \&
Ruffert 1998).  Similar arguments hold for accretion induced collapse
of white dwarfs to neutron stars and for the merger of binary white
dwarfs to neutron stars.  In fact, recent X-ray and radio observations
of supernova remnants have identified several young, isolated, rapidly
rotating pulsars, suggesting that these stars may have been born with
periods of several milliseconds (Marshall et al.~1998; Kaspi et
al.~1998; Torii et at.~1999). These neutron stars could be the
collapsed remnants of rapidly rotating progenitors.

Rapidly rotating neutron stars may naturally arise in the merger of binary
neutron stars.  Baumgarte, Shapiro and Shibata (2000) have studied
equilibrium configurations of differentially rotating neutron stars
and found examples where the maximum allowed mass increases by a factor
of about 2 due to differential rotation. This suggests that the merger of
binary neutron stars could result in a ``hypermassive'' neutron star
which has rest mass exceeding the maximum value for uniformly rotating
stars.  Recent hydrodynamic simulations in full general relativity
indicate that such hypermassive neutron stars can indeed be produced
in the merger of moderately compact neutron stars (Shibata \& Uryu
2000).  They show that the remnant is unlikely to exceed the onset
point of dynamical instability initially.  Subsequent neutrino
emission and cooling, however, will make the star shrink in size,
leading to an increase in $\beta$, possibly beyond the onset of
nonradial dynamical instability, $\beta_d$.

Rapidly rotating neutron stars experiencing the bar-mode instability
could have significant observable consequences. According to 
Newtonian simulations (Tohline, Durisen \& McCollough 1985; Durisen
Gingold \& Tohline 1986; Williams \& Tohline 1988; Houser, Centrella
\& Smith 1994; Smith, Houser \& Centrella 1995; Houser \& Centrella
1996), a dynamically unstable star may evolve into a two-component
system containing a central star and circumstellar accretion disk.
Such a system may be observable in a supernova remnant.  In the case
of merged binaries, the differentially rotating remnant may be more
massive, hot and bloated than a typical rapidly rotating, old
pulsar. Consequently, the frequency of gravitational waves excited by
the bar-mode instability could be low, {\it i.e.}, less than 1kHz (see
Eq.~(\ref{wavef}) below), and hence detectable by kilometer-size laser
interferometers such as LIGO (Lai \& Shapiro 1995; Thorne 1995).

In this paper we summarize the results of our fully relativistic
simulations of bar-mode instabilities in neutron stars.  We determine
$\beta_d$ for highly relativistic stars, follow the growth of the
bar-mode, and find the frequency and amplitude of the emitted
gravitational waves.  We implement the numerical scheme described
in Shibata (1999a), using differentially rotating neutron stars of high
$\beta$ for initial data.  We focus on differentially rotating stars,
since uniformly rotating stars do not reach $\beta \gtrsim 0.2$ except
for extremely stiff equations of state, and hence do not become
dynamically unstable to bar-modes (Tassoul 1978).  We adopt an
adiabatic equation of state with $\Gamma=2$ as a
reasonable qualitative approximation to moderately stiff nuclear
equation of state. The adiabatic assumption is justified even for hot
neutron stars, since energy dissipation is small over the dynamical
timescales of interest.

In Sec.~2, we briefly summarize our formulation of the fully
relativistic system of equations and our numerical scheme.  In Sec.~3,
initial models of differentially rotating, equilibrium neutron stars
are presented. Following Shibata, Baumgarte \& Shapiro (2000) we adopt
the so-called conformal flatness approximation to prepare
differentially rotating neutron stars in (approximate) equilibrium
states for computational convenience.  To confirm the reliability of
this approximation, we also compute numerically exact equilibrium
states and demonstrate that this approximation is accurate (cf.~Cook,
Shapiro \& Teukolsky 1996). In Sec.~4, we present our numerical
results, focusing on the onset of the bar-mode instability, its early
growth and corresponding waveforms and frequencies.  We briefly
summarize our results in Sec.~5.

Throughout this paper, we adopt geometrized units with $G=1=c$ where
$G$ and $c$ denote the gravitational constant and speed of light.  
In numerical simulation, we
use Cartesian coordinates $x^k=(x, y, z)$ with $r=\sqrt{x^2+y^2+z^2}$,
$\varpi=\sqrt{x^2+y^2}$ and $\varphi=\tan^{-1}(y/x)$; $t$ denotes
coordinate time. Greek indices $\mu, \nu, \dots$ denote $x, y, z$ and
$t$, and Latin indices $i,j,k, \dots$ denote $x,y$ and $z$.

\section{SUMMARY OF THE FORMULATION}

We perform hydrodynamic simulations in full $3+1$ general relativity
(GR).  We use the same formulation and gauge conditions as in Shibata
(1999a), to which the reader
may refer for details and basic equations.  The fundamental variables
used in this paper are:
\beqn 
\rho &&:{\rm rest~ mass~ density},\nonumber \\ 
\varep &&: {\rm specific~ internal~ energy}, \nonumber 
\\ P &&:{\rm pressure}, \nonumber \\ 
u^{\mu} &&: {\rm four~ velocity}, \nonumber \\ 
v^{k}&& ={u^k \over u^0}; ~~\Omega =v^{\varphi}, \nonumber \\ 
\alpha &&: {\rm lapse~ function}, \nonumber \\ 
\beta^k &&: {\rm shift~ vector}, \nonumber \\ 
\gamma_{ij} &&:{\rm metric~ in~ 3D~ spatial~ hypersurface},\nonumber \\ 
\gamma &&=e^{12\phi}={\rm det}(\gamma_{ij}), \nonumber \\ 
\tilde \gamma_{ij}&&=e^{-4\phi}\gamma_{ij}, \nonumber \\ 
K_{ij} &&:{\rm extrinsic~curvature}.\nonumber 
\eeqn 
Geometric variables, $\phi$, $\tilde \gamma_{ij}$, 
the trace of the extrinsic curvature $K\equiv K_{ij}\gamma^{ij}$, 
$\tilde A_{ij}\equiv e^{-4\phi}(K_{ij}-\gamma_{ij}K/3)$, 
as well as three auxiliary functions $F_i \equiv
\pa_j \tilde \gamma_{ij}$, where $\pa_j$ is the partial derivative, 
are evolved with an unconstrained evolution code in a modified form of
the ADM formalism (Shibata \& Nakamura 1995).  GR hydrodynamic
equations are evolved using a van Leer scheme for the advection terms
(van Leer 1977; Hawley, Smarr \& Wilson 1984).  
Numerical simulation is performed using 
Cartesian coordinates. Violations of the
Hamiltonian constraint and conservation of mass and angular momentum
are monitored as code checks.  Several test calculations, including
spherical collapse of dust, stability of spherical neutron stars, and
the stable evolutions of rigidly and rapidly rotating neutron stars
have been described in Shibata (1999a).  Simulations using this code and
exploring the dynamical (quasi-radial) stability against gravitational
collapse of rigidly rotating ``supramassive'' neutron stars, which
have rest masses exceeding the TOV limit for a nonrotating spherical
star, have been presented in Shibata, Baumgarte \& Shapiro
(2000). A simulation using this code and demonstrating the existence of
dynamically stable differentially rotating ``hypermassive'' stars,
which have rest masses exceeding the maximum value for uniformly
rotating stars, was presented in Baumgarte, Shapiro and Shibata
(2000).

The stress energy tensor for an ideal fluid is given by 
\beq
T_{\mu\nu}=(\rho + \rho \varep + P)u_{\mu} u_{\nu} + P g_{\mu\nu},
\eeq 
where $g_{\mu\nu}$ is the spacetime metric.  We adopt a
$\Gamma$-law equation of state
\beq 
P=(\Gamma-1)\rho \varep,\label{EOS} 
\eeq 
where $\Gamma$ is the adiabatic constant.  For isentropic configurations
the $\Gamma$-law equation of state can be rewritten in the polytropic
form 
\beq P = \kappa \rho^{\Gamma}, \mbox{~~~~~} \Gamma = 1 + \frac{1}{n}
\label{eos}, 
\eeq 
where $\kappa$ is the polytropic constant and $n$ the polytropic
index.  This is the form that we use for constructing initial data.
Throughout this paper, we adopt $n = 1$ as a reasonable qualitative
approximation to a moderately stiff, nuclear equation of state
for simplicity.

Instead of $\rho$ and $\varep$ we numerically evolve the densities
$\rho_* \equiv \rho \alpha u^0 e^{6\phi}$ and $e_* \equiv
(\rho\varepsilon)^{1/\Gamma}\alpha u^0 e^{6\phi}$ as the hydrodynamic
variables (Shibata, Oohara \& Nakamura 1997; Shibata 1999a). Since these 
variables satisfy evolution equations in conservation form, 
the total rest mass of the system 
\beq M_0=\int d^3 x \rho_*.  
\eeq 
is automatically
conserved, as is the the volume integral of the energy density $e_*$ in
the absence of shocks.

The time slicing and spatial gauge conditions we use in this paper for the
lapse and shift are the
same as those adopted in our series of papers (Shibata 1999a, 1999b;
Shibata, Baumgarte \& Shapiro 2000); i.e.~we impose an ``approximate''
maximal slice condition ($K \simeq 0$) and an ``approximate''
minimum distortion gauge condition ($\tilde D_i (\pa_t \tilde
\gamma^{ij}) \simeq 0$ where $\tilde D_i$ is the covariant derivative
with respect to $\tilde \gamma_{ij}$, see Shibata 1999b).

\section{INITIAL CONDITIONS FOR ROTATING NEUTRON STARS} 

As initial conditions, we adopt rapidly and differentially rotating
neutron stars in (approximate) equilibrium states.  The approximate
equilibrium states are obtained by choosing a conformally flat spatial
metric, {\it i.e.}, assuming $\gamma_{ij}=e^{4\phi} \delta_{ij}$ (see,
e.g., Cook, Shapiro \& Teukolsky, 1996, or Shibata 1999a for the
equations to be solved in this approximate framework).  This approach
is computationally convenient and, as demonstrated in Cook, Shapiro \&
Teukolsky (1996) and Shibata, Baumgarte \& Shapiro (2000) provides an
excellent approximation to exact axisymmetric equilibrium
configurations in rigid rotation.

Following previous studies (e.g.~Komatsu, Eriguchi \& Hachisu 1989a, 1989b;
Cook, Shapiro \& Teukolsky 1992, 1994; Bonazzola, Gourgoulhon, Salgado
\& Marck 1993; Salgado, Bonazzola, Gourgoulhon, \& Haensel 1994;
Goussard, Haensel \& Zdunik 1998) we fix the differentially rotational
profile according to
\beq 
F(\Omega) \equiv u^0
u_{\varphi}=A^2(\Omega_0 - \Omega), 
\eeq 
where $A$ is an arbitrary
constant with dimensions of length (which describes the length scale
over which $\Omega$ changes) and $\Omega_0$ is the angular velocity on
the rotation axis, which is chosen to be the $z$-axis.  In the
Newtonian limit $u^0 \rightarrow 1$ and $u_{\varphi}\rightarrow
\varpi^2 \Omega$, the rotational profile reduces to 
\beq 
\Omega={A^2 \Omega_0 \over \varpi^2 + A^2}. \label{NewOmega} 
\eeq 
Thus, for
smaller $A$, $\Omega$ becomes a steeper function of $\varpi$.

Equilibrium configurations of rotating stars are characterized by
their gravitational mass $M$, angular momentum $J$, rotational kinetic
energy $T$ and gravitational potential binding energy $W$, which in GR
can be defined invariantly according to
\beqn
M&&=\int (-2T_0^{~0}+T_{\mu}^{~\mu})\alpha e^{6\phi} d^3x,\\
J&&=\int T_{\varphi}^{~0}\alpha e^{6\phi} d^3x,\\
T&&={1 \over 2}\int \Omega T_{\varphi}^{~0}\alpha e^{6\phi} d^3x,\\
W&&=\int \rho_* \varep d^3x+T+M_0-M.
\eeqn
(e.g.~Cook, Shapiro \& Teukolsky 1992). For approximate 
configurations derived in the conformal flatness approximation, 
on the other hand, the gravitational mass is computed from the 
asymptotic behavior of the conformal factor, which, after using 
Gauss' law and the Hamiltonian constraint yields
\beq
M=\int \biggl[(\rho+\rho\varepsilon+P)(\alpha u^0)^2-P 
+ {1 \over 16\pi}K^{ij}K_{ij}\biggr] \alpha e^{5\phi} d^3x
\eeq
(see, e.g., Bowen \& York 1980).  This expression is correct
independent of axisymmetry.  For conformally flat configurations,
$J$, $T$, and $W$ can be computed from Eqs.~(8)--(10) for
nonaxisymmetric configurations as well.  As in Newtonian gravity, we
define $\beta$ as the ratio $T/W$.  (Note $W>0$ in our definition.)

Physical units enter the problem only through the polytropic constant
$\kappa$, which can be chosen arbitrarily or else completely scaled out of
the problem.  In the following, we define
\beqn \label{rescale}
&&\bar M_0 = M_0 \kappa^{-n/2}, ~~\bar M =  M \kappa^{-n/2},  \nonumber \\
&&\bar J = J \kappa^{-n},  ~~\bar P_{\rm rot} = P_{\rm rot} \kappa^{-n/2}, 
\nonumber \\
&&\bar \rho_{\rm max} = \rho_{\rm max} \kappa^n
\eeqn
where $P_{\rm rot}$ and $\rho_{\rm max}$ are rotational period and the
maximum density.  Note that $\rho_{\rm max}$ does not necessarily
coincide with the central density for stars of highly differential
rotation.  The barred quantities are now independent of $\kappa$, and
all results can be scaled for arbitrary $\kappa$ using
Eqs.~(\ref{rescale}).

For the construction of (approximate) equilibrium models for initial
data, we adopt a grid in which the semi-major axes of the stars,
chosen along the $x$ and $y$-axes in the equatorial plane, are covered
with 40 grid points.  The semi-minor axis in the polar direction along
$z$-axis is covered with $\sim 10$ grid points for the case $\beta
\sim 0.25$ and $A \sim r_e$.  Hereafter, the coordinate lengths of the
semi-major and minor axes are referred as $r_e$ and $r_p$,
respectively.  We have confirmed the convergence of 
our numerical solutions by increasing the number of grid points
covering $r_e$ to 120 for typical models 
shown in Table I.  Comparing with these higher resolution models
shows that the numerical error in $M$, $T$, $W$ and $J$ of the 
lower resolution models is less than 1 \%.

We find that for $n=1$, stars with $\beta \gtrsim 0.2$ can only be
constructed for $A \lesssim r_e$. Also, for $A \lesssim r_e/2$, stars
with $\beta \gtrsim 0.25$ are not spheroid but toroids.  Focusing on
spheroidal stars with $\beta \gtrsim 0.2$, we study cases with $\hat A
\equiv A/r_e = 0.8$ and 1.

In Table I, we display parameters for selected models constructed both
within the conformal flatness approximation and from the exact
numerical equations (using the code from Cook, Shapiro \& Teukolsky
1994) for a given $\bar \rho_{\rm max}$, $\hat A$ and $r_p/r_e$.  Note
that both approaches lead to valid, fully relativistic initial data in
the sense that both data satisfy the constraint equations of
Einstein's field equations.  The difference is that the ``exact''
solutions provide an exactly stationary solution, while the
conformally flat solutions may only be approximately stationary when
evolved dynamically.  Here, we choose stars of $R/M \sim 7$ and 9.5
(Hereafter, $R$ denotes the circumferential radius of the equator).
All the stars in Table I are ``hypermassive'' (Baumgarte, Shapiro \&
Shibata 2000) with rest masses larger than the maximum rest mass of
rigidly rotating stars built from the same equation of state ($\bar
M_0\simeq 0.207$; cf.~Fig.~2).  The matter profiles of the
equilibrium configurations obtained in the conformal flatness
approximation agree fairly well with the exact numerical solution.
The slight deviation arises mainly from the error associated with the
conformal flatness approximation (as opposed to errors due to the
different finite differencing in the two different codes).  As can be
seen in Table I, $\beta$ is systematically underestimated in the
conformal flatness approximation by $\lesssim 0.004$ ({\it i.e.}, 
$\lesssim 2\%$) although the deviation for $M_0$ and $M$ is considerably less
($< 1\%$ error).  From a post-Newtonian point of view, $T$ and
$W$ are quantities of $O(c^{-2})$ but $M_0$ and $M$ are of $O(c^0)$.
In the conformal flatness approximation, we neglect the second
post-Newtonian terms of $O(c^{-4})$ in the metric (cf.~Kley \&
Sch\"afer 1999) so that the error for $T$ and $W$ can be larger than
that for masses by $O(c^2)$
\footnote{In the conformal flatness approximation, the error in $T$
and $W$ is $O(c^{-4})$, and $O(c^{-6})$ in $M$ and $M_0$ (see, e.g.,
Asada and Shibata, 1996).  Hence, the magnitude of the error for $T$
and $W$ is expected to be $\sim (GM/Rc^2)^2 \sim 2\%$ for $R/M=7$, but
that for $M$ and $M_0$ is $\sim (GM/Rc^2)^3 < 1\%$, which is
consistent with our numerical results.}.  Indeed, the agreement
between exact and approximate solutions is improved for less compact
stars (compare models D2 and D3 with D6 and D7).  Because of this
small deviation from the exact solution, the initial conditions
prepared in the conformal flatness approximation should be regarded as
slightly perturbed states of exact equilibria.

In Fig. 1, we plot $\Omega/\Omega_0$ as a function of $\varpi$ in the
equatorial plane to illustrate the rotational velocity field for
different $\hat A$.  We show an unstable (to bar-modes) star with $R/M
\sim 7$ and $\beta \sim 0.25$, for both $\hat A = 1$ and $0.8$. As
expected from Eq.~(\ref{NewOmega}), $\Omega$ is a steeper function of
$\varpi$ for smaller $\hat A$. As demonstrated in the Newtonian
calculations of Pickett, Durisen and Davis (1996), the degree of
falloff of $\Omega$ versus $\varpi$ could be an important factor for
determining the onset point of nonaxisymmetric dynamical instability
in differentially rotating stars.

\section{NUMERICAL RESULTS}

To investigate the dynamical stability against bar-mode
deformation, we initially superimpose a density perturbation of the form 
\beq
\rho=\rho_{0}\Bigl(1+\delta_b {x^2-y^2 \over r_e^2}\Bigr), 
\eeq 
where $\rho_0$ denotes the axisymmetric configuration, and we choose
$\delta_b$ to be 0.1 or 0.3.  We leave the four-velocity
$u_i$ unperturbed.  We recompute the constraint (initial value)
equations whenever we modify the equilibrium configurations this way,
to guarantee that we are satisfying the Einstein equations at $t=0$.

The growth of a bar mode can be followed by monitoring the distortion
parameter
\beq
\eta \equiv 2{x_{\rm rms}-y_{\rm rms} \over x_{\rm rms}+y_{\rm rms}},
\eeq
where $x^i_{\rm rms}$ denotes the mean square axial length 
\beq
x^i_{\rm rms}=\biggl[
{1 \over M_0}\int d^3x \rho_*~(x^i)^2\biggr]^{1/2}.
\eeq
For dynamically unstable stars, $\eta$ grows exponentially until 
reaching a saturation point, while for stable stars, it remains
approximately constant for many rotational periods.

We perform simulations using a fixed uniform grid with typical size
$153\times 77 \times 77$ in $x-y-z$ 
and assume $\pi$-rotation symmetry
around the $z$-axis as well as a reflection symmetry about
the $z=0$ plane.  We have also performed test simulations with
different grid resolutions to check that the results do not change
significantly.  Since we impose $\pi$-rotation symmetry, we ignore
one-armed spiral ($m=1$) modes which might be dominant for rotating
stars in which $\Omega$ is a very steep function of
$\varpi$ (Pickett, Durisen \& Davis 1996). However, $\pi$-rotation 
symmetry guarantees that the center-of-mass drift is identically 
zero (cf.~New, Centrella \& Tohline 1999). 

We note that the outer boundaries of our computational domain reside
inside the wavelength of gravitational waves emitted by the bar-mode
perturbation, $\lambda_{\rm gw}$.  The typical location of the outer
boundaries along each axis is $\sim (0.1-0.2) \lambda_{\rm gw}$. Without
setting the boundaries in the radiation zone at $r \gtrsim
\lambda_{\rm gw}$, or using a sophisticated wave-extraction technique
in the near zone (Bishop et al.~1996; 
Abrahams et al.~1998), it is impossible to compute the
radiation reaction and asymptotic waveforms completely accurately.
However, in this paper we focus on dynamical instabilities, which are
independent of dissipation processes and grow on a dynamical timescale
considerably shorter than the secular dissipation timescale due to
gravitational wave emission.  The error in the evaluation of the
gravitational waves can therefore be safely neglected in assessing the
onset and growth of a dynamical instability.

Fig.~2 summarizes our findings on the dynamical stability against
bar-mode deformation.  We evolve a range of stellar models with $\hat
A = 0.8$ (squares) and $\hat A = 1$ (circles).  All of these models
have $\beta \geq 0.2$ and $0.2 < r_p/r_e < 0.35$, and we determine
their stability by inducing an initial nonaxisymmetric perturbation
$\delta_b=0.1$.  Each model takes 50 -- 100 CPU hours to run on the
FACOM VX/4R machine; the long-time runs described below take about 150
CPU hours.  Stable stars are denoted with open circles and squares,
and unstable stars with solid circles or squares.  For models denoted
by crosses we were unable to determine stability unambiguously.  Note
that all differentially rotating stars shown here are dynamically
stable against quasi-radial collapse to black holes.

Fig.~2 shows that in a $\bar M_0$ versus $\bar \rho_{\rm max}$ diagram
an unstable region can be clearly separated from a stable region. The
demarcation line is nearly independent of the degree of differential
rotation, at least for the modest variation in the rotation law that
we consider (recall that for $\beta \gtrsim 0.2$ spheroidal stars only
exist within a restricted range of $\hat A$).  The two regions are
separated by a thick dashed-dotted line in the figure.  We have also
plotted lines of constant $\beta$; one where $\beta=0.245$ for $\hat
A=1$ and another where $\beta=0.24$ for $\hat A=0.8$.  These two lines
very closely trace the demarcation line between the regions of
stability and instability.  This result suggests that {\it in general
relativity as in Newtonian gravitation, the parameter $\beta$ is a
good diagnostic for assessing whether a rotating star is stable
against the dynamical bar mode instability.} It also suggests that for
differentially rotating, relativistic stellar models, the threshold
for dynamical bar formation $\beta_{dGR}$ may depend only weakly on
the differential rotation law, and is only slightly smaller than the
corresponding value for uniformly rotating, Newtonian stars, $\beta_d
\sim 0.27$.

The small but measurable decrease in the critical value of $\beta$
could be either due to the presence of differential rotation
(cf.~Tohline \& Hachisu (1990) and Pickett, Durisen \& Davis (1996),
who have observed this effect in Newtonian gravity), or by GR effects
(compare with the decrease in $\beta_s$ with increasing compaction for
the secular onset of the gravitational-wave driven instability in
Stergioulas \& Friedman 1998), or a combination of both.  In order to
separate the two effects it would be desirable to systematically
explore parameter space and study models with different rotation laws
and varying compaction up to the Newtonian limit of small $M/R$. We
are preparing such a survey now (Saijo, Shibata, Baumgarte \& Shapiro
2000).

In anticipation of this survey, recall that in Newtonian gravity,
$\bar M_0$ scales with $\rho_{\rm max}^{(3-n)/2n}$ (Shapiro \&
Teukolsky 1983) for polytropes, or $\bar M_0 \propto \bar \rho_{\rm
max}$ for $n=1$.  For stars with $\beta \simeq 0.27$, this relation
turns out to be $\bar M_0 \simeq 10 - 12 \bar \rho_{\rm max}$; for smaller
values of $\beta$ the coefficient is only slightly smaller.  We
therefore expect that the line marking the onset of dynamical
instability, a line of constant $\beta$, approaches a linear
relationship in Fig.~2 in the Newtonian limit.

Unfortunately, using a fully relativistic code is impractical for
simulating stars in the Newtonian or post-Newtonian regime.  The
Courant condition restricts the numerical timestep to the light travel
time across a grid zone, and therefore scales with $R$.  The dynamical
timescale of the star, however, is approximately the free-fall
timescale $R^{3/2}/M^{1/2}$, so that the ratio between the dynamical
timescale and Courant timestep scales roughly as $(R/M)^{1/2}$.  In
the Newtonian limit this ratio becomes very large, so that many
timesteps have to be carried out in order to simulate a fixed number
of dynamical timescales, which makes the calculation computationally
impractical.  To avoid this problem, we are implementing a
post-Newtonian code (Shibata, Baumgarte \& Shapiro 1998) to explore
the intermediary regime, and will present these results in a
forthcoming paper (Saijo, Shibata, Baumgarte \& Shapiro 2000). Our
preliminary finding is that in the Newtonian limit, $\beta_d \sim
0.26$ for $\hat A = 1$, and hence the onset of dynamical instability
occurs slightly earlier for stars of greater compaction.

For five of the above models (D1, D2, D3, D6 and D7) we have followed
the growth and evolution of the bar-mode instability over several
rotational timescales.  In Figs.~3, 4 and 5 we show snapshots of
density contours and velocity fields in the $x-y$ (left) and $x-z$
(right) planes for the compact models D1, D2 and D3 ($R/M \sim 7$; see
Table I).  The models are rotating counterclockwise.  Note that the
stars start out as highly flattened, disk-like objects.  In Fig.~6,
we also show $\eta$ as a function of time (a) for models D1, D2 and D3
as well as (b) for models D6 and D7, which are slightly less compact ($R/M
\sim 9.5$).  
In order to accelerate the growth of the
instabilities, we set $\delta_b=0.3$ for these simulations.
In addition, we plot the early evolution of $|\eta|$ on a semi-log scale 
in Fig.~7.  This plot also demonstrates that while the perturbation parameter
$\delta_b=0.3$ is fairly large, the nondimensional measure $|\eta|$
is initially safely in the linear regime.
Unstable
growth ceases once the bar-mode reaches nonlinear saturation.
Following the evolution much beyond this point is impractical, both
because of accumulation of numerical error, and because further
evolution begins to be affected by gravitational wave emission, which
is crudely handled in this code as discussed above.

Model D1 is stable against bar-mode formation and demonstrates the
ability of our code to identify and maintain such a configuration in
stable equilibrium (see Fig.~3).  The other four models are unstable
and the barlike perturbation (and hence $\eta$) grows exponentially in
the early phase until saturation is reached, typically when $\eta \sim
0.2-0.4$.  However, the evolution after the saturation varies for
different models. For model D2, spiral arms form in the outer part of
the barlike object, which then spread outward, transporting away mass
and angular momentum (see Fig.~4).  In Fig. 8, we show the fraction of
the rest mass inside a fixed coordinate radius $r$, $M_*(r)/M_*$, as a
function of time for model D2. We find that a few per cent of the
total rest mass is ejected from the star.  We also find that the
fraction of the rest mass inside $r=0.2r_e$ and $0.6r_e$ ultimately
{\it increases} during the late phase of the evolution since the mass
with high specific angular momentum is transported outwards, and the
star slightly contracts. As a consequence, the maximum density of the
star increases at late times by about $50\%$ from $\bar \rho_{\rm max}
\simeq 0.045$ to $\sim 0.065$ by $t=6P_{\rm rot}^a$\footnote{For the
stable model D1 we find a small increases in the central density as
well, but only by about 10 \% after $t=6P_{\rm rot}^a$.  This small
increase is due to numerical viscosity and is a numerical artifact.
The central density in the model D2 increases by a much larger
fraction over the same timescale, suggesting that the bulk of this
increase is indeed a physical effect.}.  Here $P_{\rm rot}^{a}$ is the
rotational period on the rotation axis ($\varpi = 0$).  Tracking its
motion in Fig.~2, we find that the unstable star approaches the
stability threshold line from the left, and ultimately enters the
stable region.  Model D6 evolves very similarly.  These findings are
also in qualitative agreement with Newtonian results (Tohline, Durisen
\& McCollough 1985; Durisen Gingold \& Tohline 1986; Williams \&
Tohline 1988; Houser, Centrella \& Smith 1994; Smith, Houser \&
Centrella 1995; Houser \& Centrella 1996; New, Centrella \& Tohline
1999).

Unstable models D3 (see Fig.~5) and D7 start out closer to the
stability threshold ($\beta \gtrsim \beta_{dGR}$) and never form
spiral arms or eject mass.  Instead, they evolve to an ellipsoidal
shape, which is maintained for many rotational periods.  Although the
mass is not ejected in this case, the maximum density slightly
increases again due to outward angular momentum
transport. Consequently, the stars again approach the threshold line
of the dynamical stability shown in Fig. 2 and become stable.

We determine the growth time and oscillation period of the bar-mode 
instability from Figs.~7 and summarize the results 
in Table II.  From Figs.~7, we can match $\eta$ to a function
\beq
\eta \simeq \eta_0 10^{t/\tau_g}\cos(2\pi t/\tau_o+\varphi_0)
\eeq
where $\eta_0$ and $\varphi_0$ are constants, and 
$\tau_g$ and $\tau_o$ are the growth time and oscillation period. 
The growth time $\tau_g$ depends strongly on the compaction 
$R/M$ and $\beta$, and is smaller for larger $\beta$ as expected. 
The oscillation period $\tau_o \simeq 1.2P_{\rm rot}^{a}$ 
depends only very weakly on the compaction $R/M$ and $\beta$, 
and approximately agrees with that found in 
Newtonian simulations for $n=1$ (Williams \& Tohline 1988).
The characteristic oscillation period of $\eta$ after 
the saturation of the growth is $\tau_o \sim (1.2 -
1.4)P_{\rm rot}^{a}$ for all models, so that the pattern period is
$\sim (2.4-2.8)P_{\rm rot}^{a}$.  

In order to check the numerical convergence of our results, we
repeated these simulations with a lower resolution ($101\times
51\times 51$ as opposed to $153\times 77\times 77$, with the outer
boundaries at the same location).  In Figs.~6 (a) and (b), the dotted
lines denote the low-resolution result for $\eta$ as a function of
time.  For $t/P_{\rm rot}^{a} \lesssim 5$ the results agree well, implying
that a fair qualitative convergence has been achieved.  For later times, the
accumulation of numerical truncation error and problems associated
with the inadequate outer boundaries results in a poorer agreement
between the two resolutions.

The differentially rotating ellipsoids formed after saturation are
highly flattened and still seem to have high $\beta \gtrsim 0.2$.  We
expect these to be secularly unstable against gravitational radiation
(Stergioulas \& Friedman 1998).  Therefore, $\eta$ will probably
maintain a fairly high value of $O(0.1)$ on a radiation reaction
timescale, which is much longer than the rotational period. As argued
by Lai \& Shapiro (1995), such an ellipsoid will ultimately settle
down to an axisymmetric star or a Dedekind-like ellipsoid (a
nonaxisymmetric, stationary star whose figure does not rotate but
which has internal differential motion, cf.~Chandrasekhar 1969).

As a measure of gravitational waveforms we show $h_+$ and $h_{\times}$
\footnote{Our quantities $h_+$ and $h_{\times}$ differ from those defined
in Misner, Thorne \& Wheeler (1973) by a factor r/M.} 
\beqn
&&h_+ \equiv r(\tilde \gamma_{xx} - \tilde \gamma_{yy})/2M\\
&&h_{\times} \equiv  r \tilde \gamma_{xy} /M
\eeqn
in Fig.~9 for models D1 (dotted lines), D2 (solid lines) and D3
(dashed lines).  These quantities are read off near the outer
boundaries along the $z$-axis as a function of retarded time.  For
models D6 and D7 very similar waveforms to those of D2 and D3 are
generated, as expected from Fig.~6.  As mentioned above, the outer
boundaries are not located in the wave zone, which implies that these
quantities will not exactly agree with the asymptotic waveforms.  We
guess that the wave shapes are in fairly good agreement with the exact
ones because the frequency agrees with the oscillation frequency of
the bar pattern, but the error in the amplitude may be $\gtrsim 10\%$
or larger.

We find that for unstable stars, the maximum magnitude of $h_+$ and
$h_{\times}$ is $\sim 0.03-0.08$ depending on $\beta$ and $R/M$, while
they remain of $O(10^{-3})$ for the stable star D1.  The maximum
observable wave amplitude from such a source situated at a distance $r$
from the earth is then approximately
\beq
h \sim 6\times 10^{-22} \biggl( {h_{+,\times} \over 0.05}\biggr)
\biggl( {M \over 2.5 M_{\odot}}\biggr)
\biggl({10 {\rm Mpc} \over r}\biggr). \label{wave}
\eeq
The typical gravitational wavelength after saturation 
is $\sim 1.3 P_{\rm rot}^{a}$ or, using the
empirical relation $P_{\rm rot}^{a} \sim 3 M (R/M)^{3/2}$ (cf.~Table
I), approximately $\sim 70 M (R/7M)^{3/2}$.  This implies that
the frequency of gravitational waves is
\beq 
f \sim 1.2 \biggl({R \over 7M}\biggr)^{-3/2} 
\biggl({2.5M_{\odot} \over M}\biggr) {\rm kHz}.
\label{wavef}
\eeq
We note that the amplitude and frequency are in approximate agreement
with earlier Newtonian calculations (Houser, Centrella \& Smith 1994;
Smith, Houser \& Centrella 1995; Houser \& Centrella 1996), which
suggests that GR effects do not drastically alter the simple
quadrupole-formula predictions for the waveforms.

Eqs.~(\ref{wave}) and (\ref{wavef}) give the maximum amplitude for one
cycle and its frequency.  As discussed in Lai \& Shapiro
(1995), the effective amplitude can be much larger because of the 
quasi-periodic nature of the source.  Also, the frequency will 
gradually shift to smaller values as a result of radiation reaction. 
Thus, even if $h$ in one cycle is small and the frequency is initially 
as high $\gtrsim 1$kHz, these nearly-ellipsoidal stars may eventually 
be observable by kilometer size laser-interferometric gravitational 
wave detectors like LIGO (Thorne 1995) for stars with $M \sim 
2.5M_{\odot}$ and $R \sim 7M$ 
as a result of the secular decrease of the frequency.  Clearly, this 
predicted drift needs to be confirmed by a more detailed study.  Also, 
gravitational waves could be a source for a specially designed narrow 
band interferometers or resonant-mass detectors in which the frequency 
band is between 1 and 2 kHz (Thorne 1995).

\section{SUMMARY}

We have performed numerical simulations of rapidly and differentially
rotating neutron stars in full $3+1$ general relativity.  We treated
compact stars of $10 \gtrsim R/M \gtrsim 6$ and focussed on their dynamical
stability against bar-mode formation.  We found that when plotted in a
$\bar M_0$ versus $\bar \rho_{\rm max}$ diagram, a region of stable
stars can be clearly distinguished from a region of unstable stars,
with the onset of instability almost independent of the degree of
differential rotation.  We showed that the parameter $\beta=T/W$
remains a good diagnostic of the onset point of instability in the
relativistic domain as it did for Newtonian stars. The critical value
for the instability onset depends only weakly on the degree of
differential rotation for the models surveyed to date.  For those
cases we find that $\beta_{dGR} \sim 0.24-0.25$, and that
$\beta_{dGR}$ decreases slightly for stars with a higher degree of
differential rotation.  We also have preliminary evidence that
$\beta_{dGR}$ decreases with compaction as well.  We will
systematically study this hypothesis with a post-Newtonian numerical
analysis in a forthcoming paper (Saijo, Shibata, Baumgarte \& Shapiro
2000).

For selected models, we followed the growth and saturation of bar-mode
perturbations up to late times.  Stars with sufficiently large $\beta
> \beta_{dGR}$ develop bars first and then form spiral arms, leading
to mass ejection.  Stars with smaller values of $\beta \sim
\beta_{dGR}$ also develop bars, but do not form spiral arms and eject
only very little mass.  In both cases, unstable stars appear to form
differentially rotating, triaxial ellipsoids once the bar-mode
perturbation saturates.  Typically, these flattened ellipsoids appear
to have $\beta \gtrsim 0.2$, so that they would be secularly unstable
due to gravitational waves and viscosity.  We expect that this secular
instability will allow the stars to maintain a bar-like shape for many
dynamical timescales, leading to quasi-periodic emission of
gravitational waves.

We estimate the initial frequency and amplitude of gravitational waves
to be $f \sim (1-1.4) $kHz and $h \sim 5\times 10^{-22}$ for stars of
mass $\sim 2.5M_{\odot}$ and radius $R \sim 7M$ at a distance of
$10$Mpc. The effective amplitude of gravitational waves could be much
larger during the subsequent evolution because of the accumulation of
quasi-periodic wave cycles (Lai \& Shapiro 1995).  In order to
accurately determine the secular evolution of the ellipsoidal star
together with emitted gravitational wave signal, a more detailed
calculation is necessary.  Since the secular timescale is larger than
the dynamical timescale by several orders of magnitude, it may be
impossible to follow the evolution with a fully dynamical code, even
with implicit differencing to avoid the Courant criterion for
stability.  This suggests that in full GR, the secular evolution
problem may best be solved within an appropriate, quasi-stationary
scheme similar in spirit to the approach used in stellar evolution
calculations.

\acknowledgments

Numerical computations were performed on the VX/4R
machines in the data processing center of the National Astronomical
Observatory of Japan.  This work was supported by NSF Grants AST
96-18524 and PHY 99-02833 and NASA Grant NAG5-7152 at the University
of Illinois at Urbana-Champaign (UIUC).  M.S. gratefully acknowledges
support by JSPS (Fellowships for Research Abroad) and the hospitality
of the Department of Physics at UIUC.

\clearpage
\onecolumn

\vskip 5mm
\noindent 
{\bf Table I.~} Maximum density $\bar \rho_{\rm max}$, 
rest mass $\bar M_0$, gravitational mass $\bar M$, 
compaction $R/M$, 
$\beta=T/W$, rotation period on the rotation axis $\varpi=0$ 
($\bar P_{\rm rot}^{a}$) 
and at the equator
($\bar P_{\rm rot}^{e}$) of differentially rotating stars 
for selected models 
computed in the conformally flatness approximation (upper line) 
and in the exact equations (lower line). 
For D1 -- D5, $R/M \sim 7$ and for D6 and D7, it is $\sim 9.5$. 
$J/M^2$ is larger than unity for the models shown here. 

\vskip 5mm
\noindent
\begin{center}
\begin{tabular}{|c|c|c|c|c|c|c|c|c|c|c|} \hline
$r_p/r_e$ & \hspace{2mm} $\hat A$ \hspace{2mm} & 
\hspace{2mm} $\bar \rho_{\rm max}$ \hspace{2mm} &
\hspace{2mm} ${\bar M_0}$ \hspace{2mm} & 
\hspace{2mm} ${\bar M}$ \hspace{2mm} &
\hspace{2mm} $R/M$      \hspace{2mm} & 
\hspace{2mm} $T/W$      \hspace{2mm} &
\hspace{2mm} $\bar P_{\rm rot}^{a}$  \hspace{2mm} &
\hspace{2mm} $\bar P_{\rm rot}^{e}$  \hspace{2mm} & stability & Model \\ \hline
0.35 & 1 & 0.06056& 0.260 &0.241 &6.62 &0.230 & 12.7 & 36.7 & stable 
&D1  \\ \hline 
     &   &        & 0.259 &0.241 &6.61 &0.233 & 12.7 &      &  &  \\ \hline
0.275& 1 & 0.04460& 0.277 &0.259 &7.07 &0.258 & 15.0 & 41.8 & unstable 
&D2  \\ \hline
     &   &        & 0.277 &0.258 &7.08 &0.262 & 14.9 &      &  &  \\ \hline
0.30 & 1 & 0.04590& 0.264 &0.246 &7.24 &0.251 & 14.9 & 41.3 & unstable 
&D3  \\ \hline 
     &   &        & 0.264 &0.246 &7.23 &0.254 & 14.8 &      &  &  \\ \hline
0.25 &0.8& 0.04650& 0.262 &0.245 &7.02 &0.243 & 12.3 & 45.3 & unstable 
&D4  \\ \hline
     &   &        & 0.261 &0.244 &7.02 &0.247 & 12.2 &      &  & \\ \hline
0.325&0.8& 0.05940& 0.254 &0.235 &6.50 &0.223 & 10.5 & 40.2 & stable 
&D5  \\ \hline
     &   &        & 0.253 &0.235 &6.50 &0.226 & 10.4 &      & & \\ \hline
0.275& 1 & 0.03070& 0.229 &0.217 &9.27 &0.262 & 19.8 & 50.7 & unstable 
&D6  \\ \hline
     &   &        & 0.229 &0.217 &9.27 &0.265 & 19.8 &      &  & \\ \hline
0.3 & 1  & 0.03220& 0.219 &0.208 &9.41 &0.254 & 19.5 & 49.7 & unstable 
&D7  \\ \hline
     &   &        & 0.219 &0.208 &9.39 &0.256 & 19.4 &      & & \\ \hline
\end{tabular}
\end{center}

\vskip 5mm
\noindent 
{\bf Table II.~} Growth time $\tau_g$ and oscillation period $\tau_o$ 
of the bar-mode 
for unstable stars D2, D3, D6, and D7. 
The timescales are shown in units of $P_{\rm rot}^{a}$.  
\vskip 5mm
\noindent
\begin{center}
\begin{tabular}{|c|c|c|} \hline
Model & \hspace{2mm} $\tau_g$ \hspace{2mm} & 
\hspace{2mm} $\tau_o$ \hspace{2mm} \\ \hline
D2 & 2.48 & 1.21 \\ \hline 
D3 & 3.62 & 1.19 \\ \hline 
D6 & 2.07 & 1.23 \\ \hline 
D7 & 2.85 & 1.21 \\ \hline 
\end{tabular}
\end{center}

\vskip 1cm

\begin{figure}[t]
\begin{center}
\epsfxsize=4in
\leavevmode
\epsffile{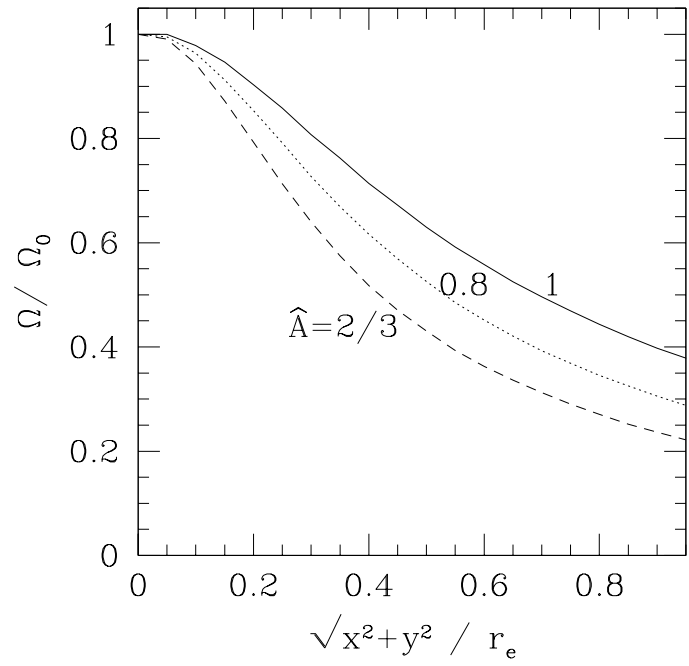}
\end{center}
\caption{$\Omega/\Omega_0$ as a function of $\varpi$ in the 
equatorial plane for differentially rotating stars of $R/M \sim 7$ 
and $\beta \sim \beta_{dGR}$ and for $\hat A=1$, 0.8 and $2/3$.}
\end{figure}

\begin{figure}[t]
\begin{center}
\epsfxsize=5.8in
\leavevmode
\epsffile{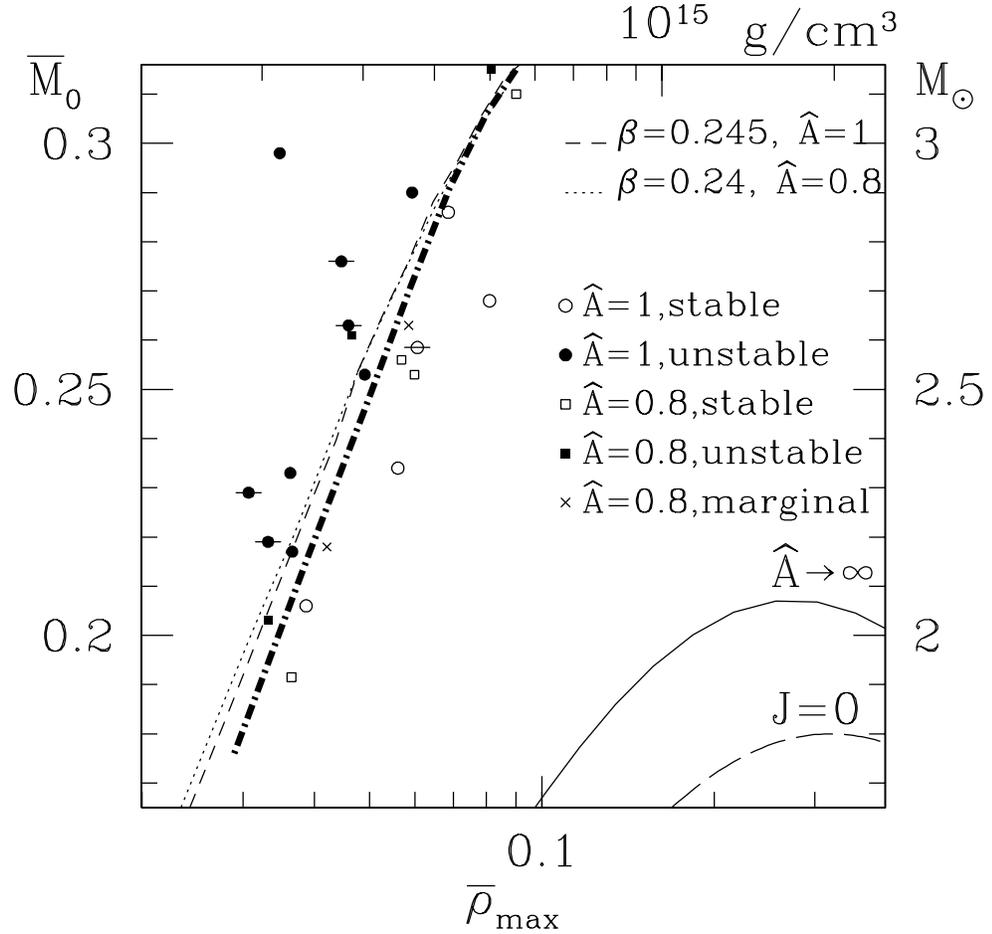}
\end{center}
\caption{Models of differentially rotating stars in a $\bar M_0$
versus $\bar \rho_{\rm max}$ diagram.  Circles denote stars with $\hat
A = 1$, squares with $\hat A = 0.8$.  Solid (open) circles or squares
represent stars that are unstable (stable).  Marginally stable stars
are denoted with a cross.  The region for the stable stars is clearly
separated from that for the unstable stars by the thick dashed-dotted
line.  This line is followed fairly closely by the dashed and dotted
lines, which have been constructed for differentially rotating stars
of $(\hat A,\beta)=(1,0.245)$ and $(0.8,0.24)$.  We carry out
long-duration simulations for those stars denoted by a hyphen (models
D1, D2, D3, D6 and D7). The long-dashed and solid lines are for
nonrotating spherical stars and rigidly rotating stars at the mass
shedding limit.  Scales for the top and right axes are shown for
$\kappa =100(G^3M_{\odot}^2/c^4)$ in which the maximum rest mass for
spherical stars is about $1.8M_{\odot}$.  }
\end{figure}

\begin{figure}[t]
\begin{center}
\epsfxsize=2.5in
\leavevmode
\epsffile{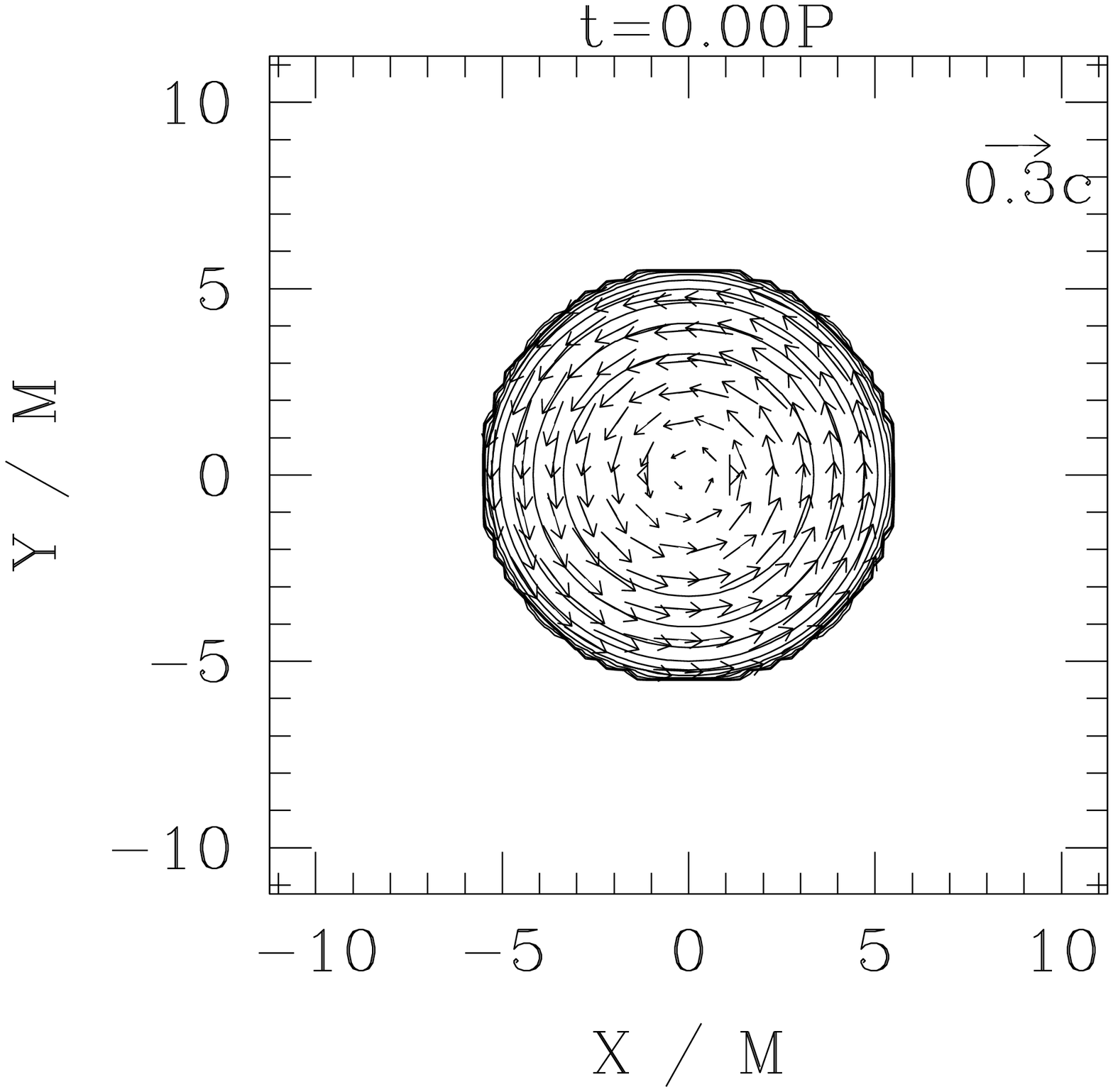}
\epsfxsize=2.5in
\leavevmode
\epsffile{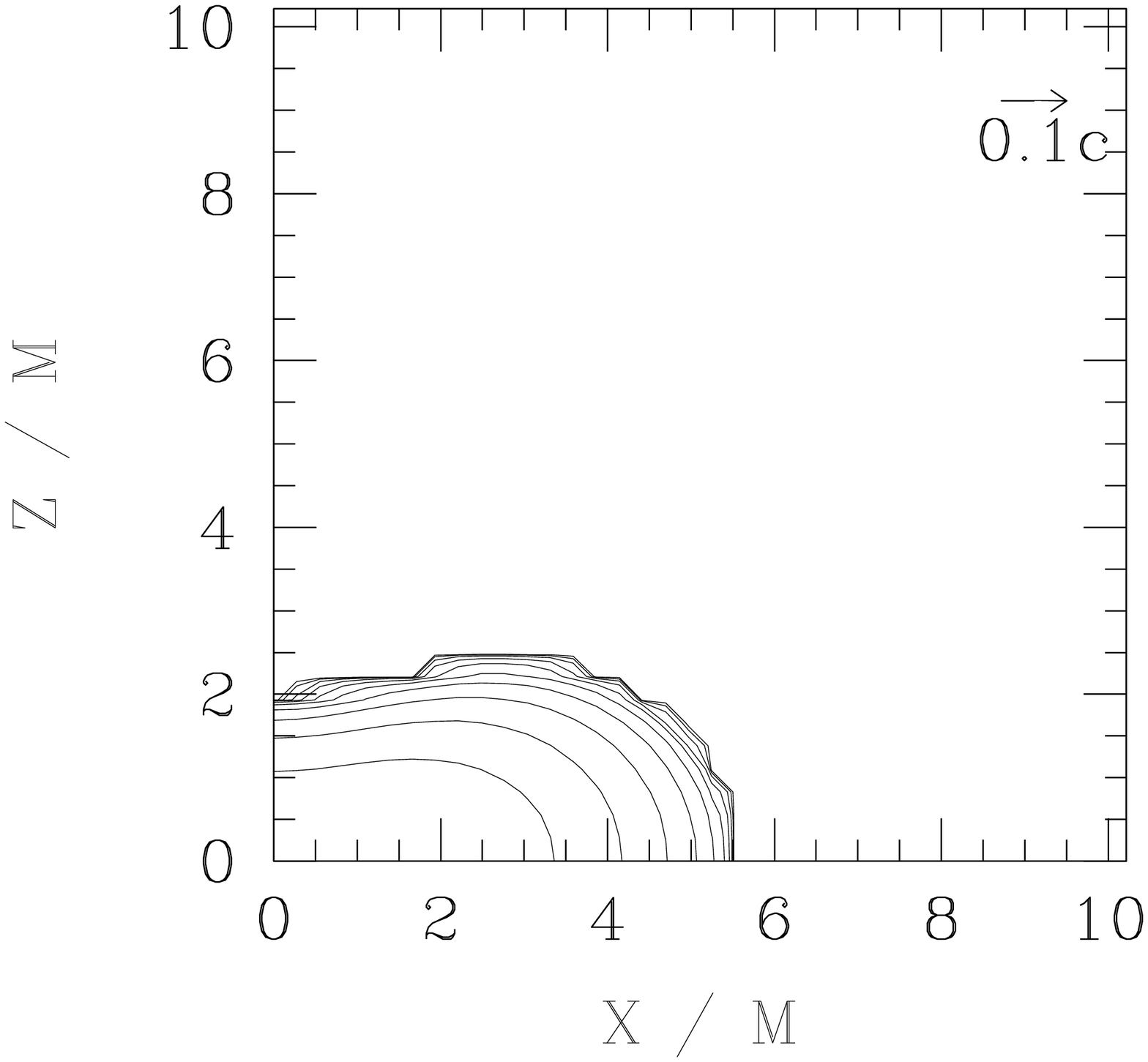}\\
\epsfxsize=2.5in
\leavevmode
\epsffile{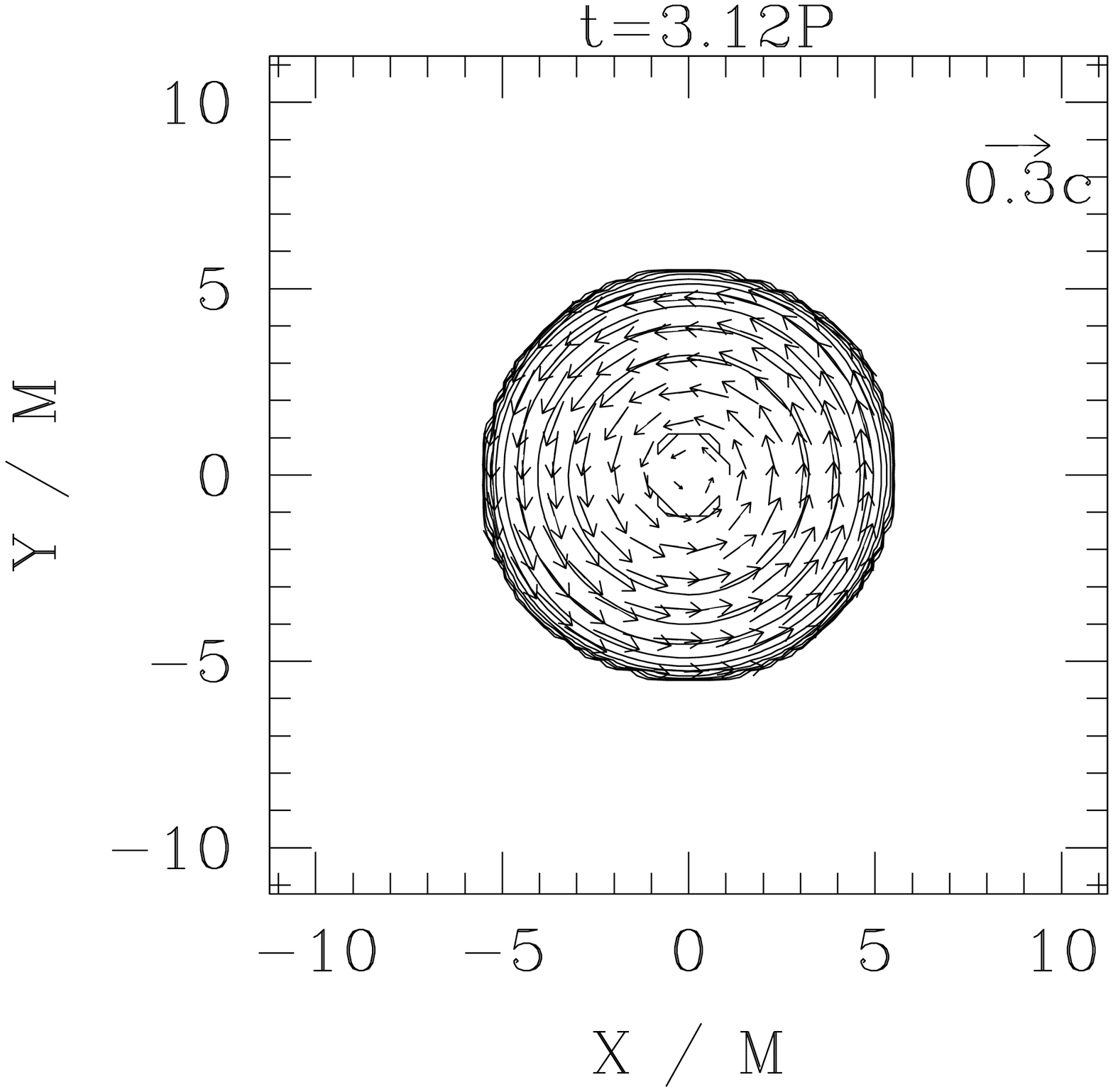}
\epsfxsize=2.5in
\leavevmode
\epsffile{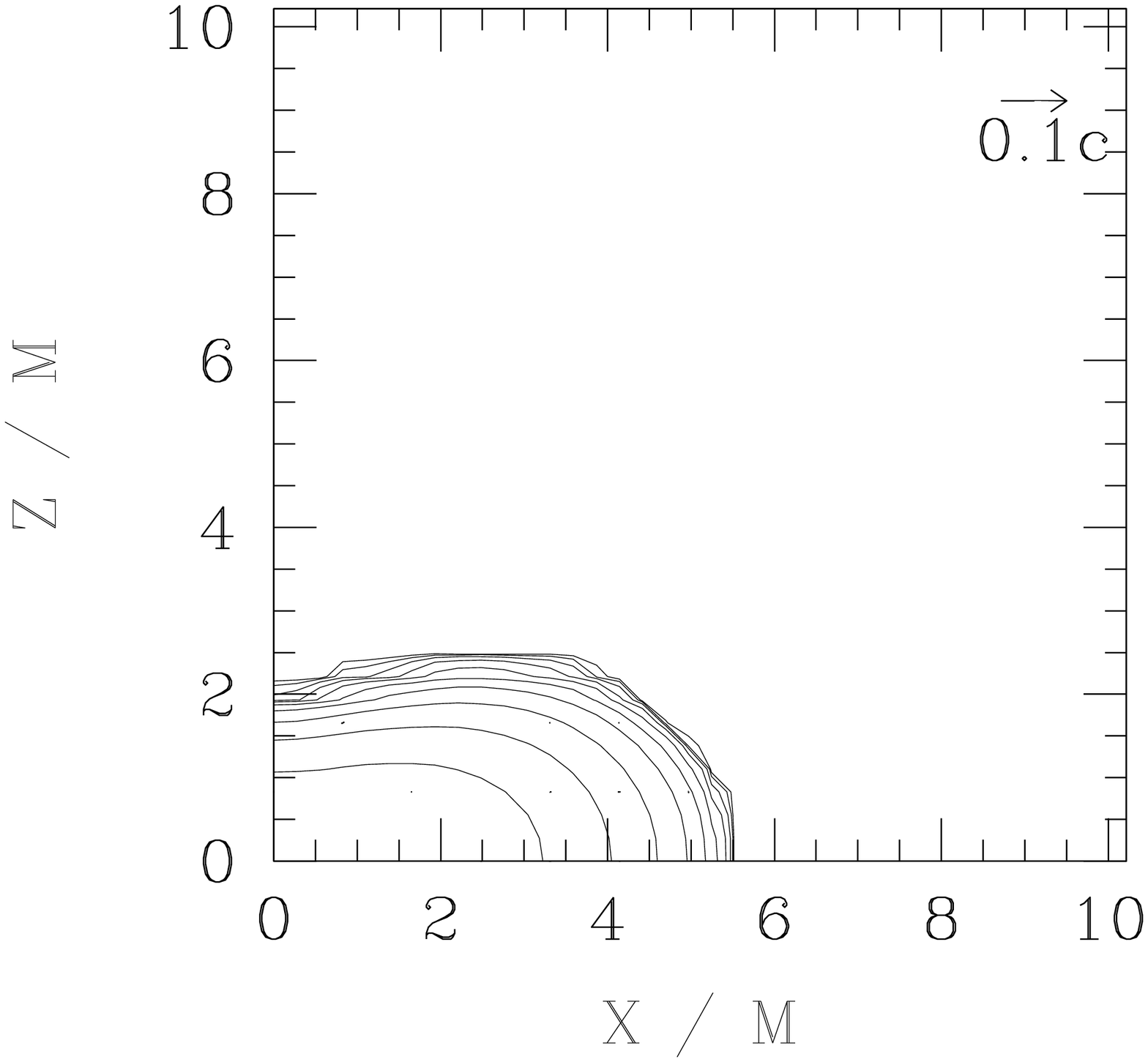}\\
\epsfxsize=2.5in
\leavevmode
\epsffile{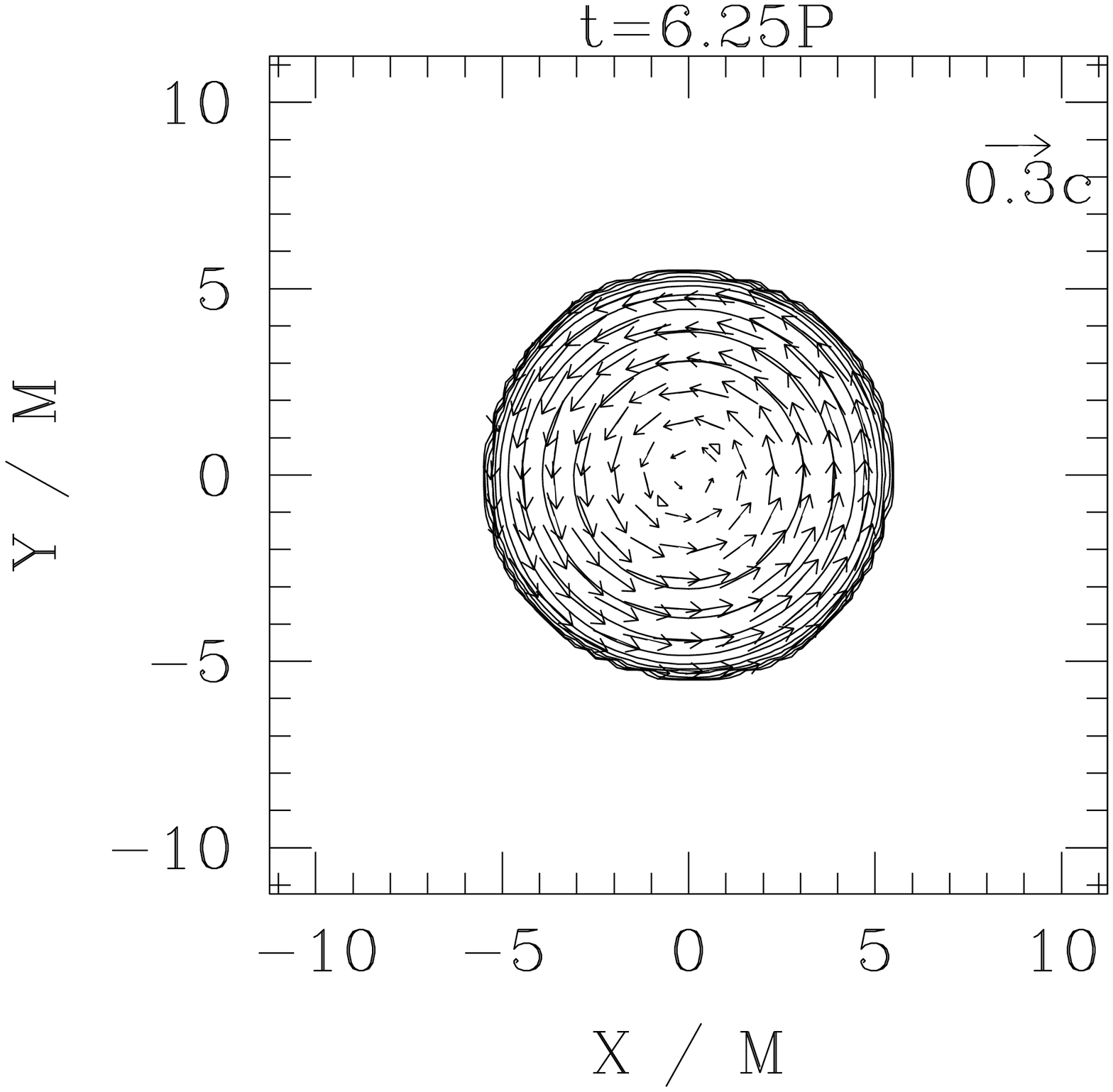}
\epsfxsize=2.5in
\leavevmode
\epsffile{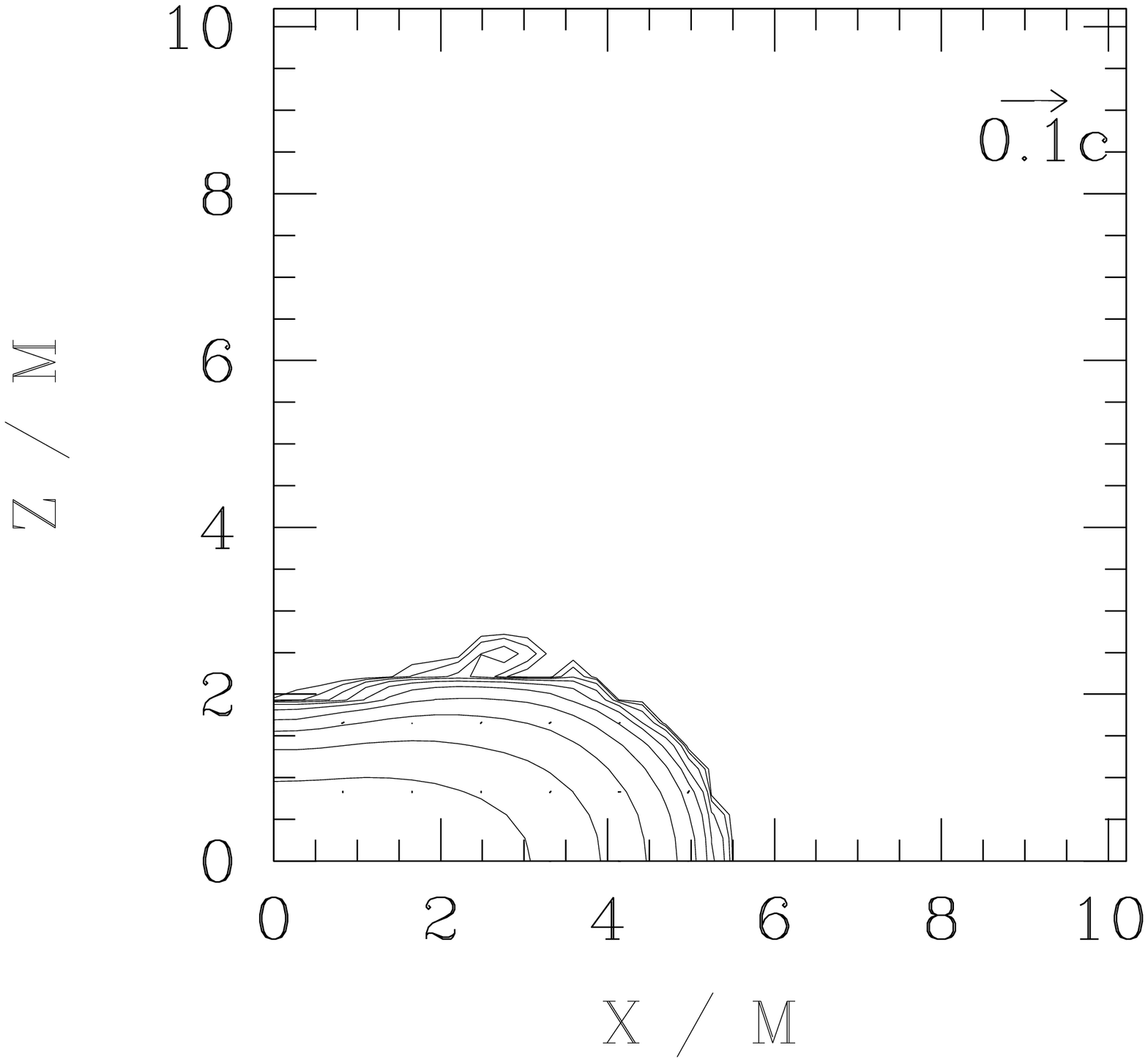}
\end{center}
\caption{Snapshots of density contours for $\rho_*$ and the velocity
flow for $v^i$ in the equatorial plane (left) and in the $y=0$ plane
(right) for the stable model D1. The contour lines are drawn for
$\rho_*/\rho_{*~{\rm max}}=10^{-0.3j}$ for $j=0,1,2,\cdots,10$ where
$\bar \rho_{*~{\rm max}}$ is 0.193, 0.209 and 0.248 for the three
different times (the corresponding values of $\bar \rho_{\rm max}$ are
0.061, 0.064, and 0.067).  The lengths of arrows are normalized to
$0.3c$ (left) and $0.1c$ (right). The time is shown in units of
$P_{\rm rot}^a$. }
\end{figure}

\clearpage
\begin{figure}[t]
\begin{center}
\epsfxsize=2.5in
\leavevmode
\epsffile{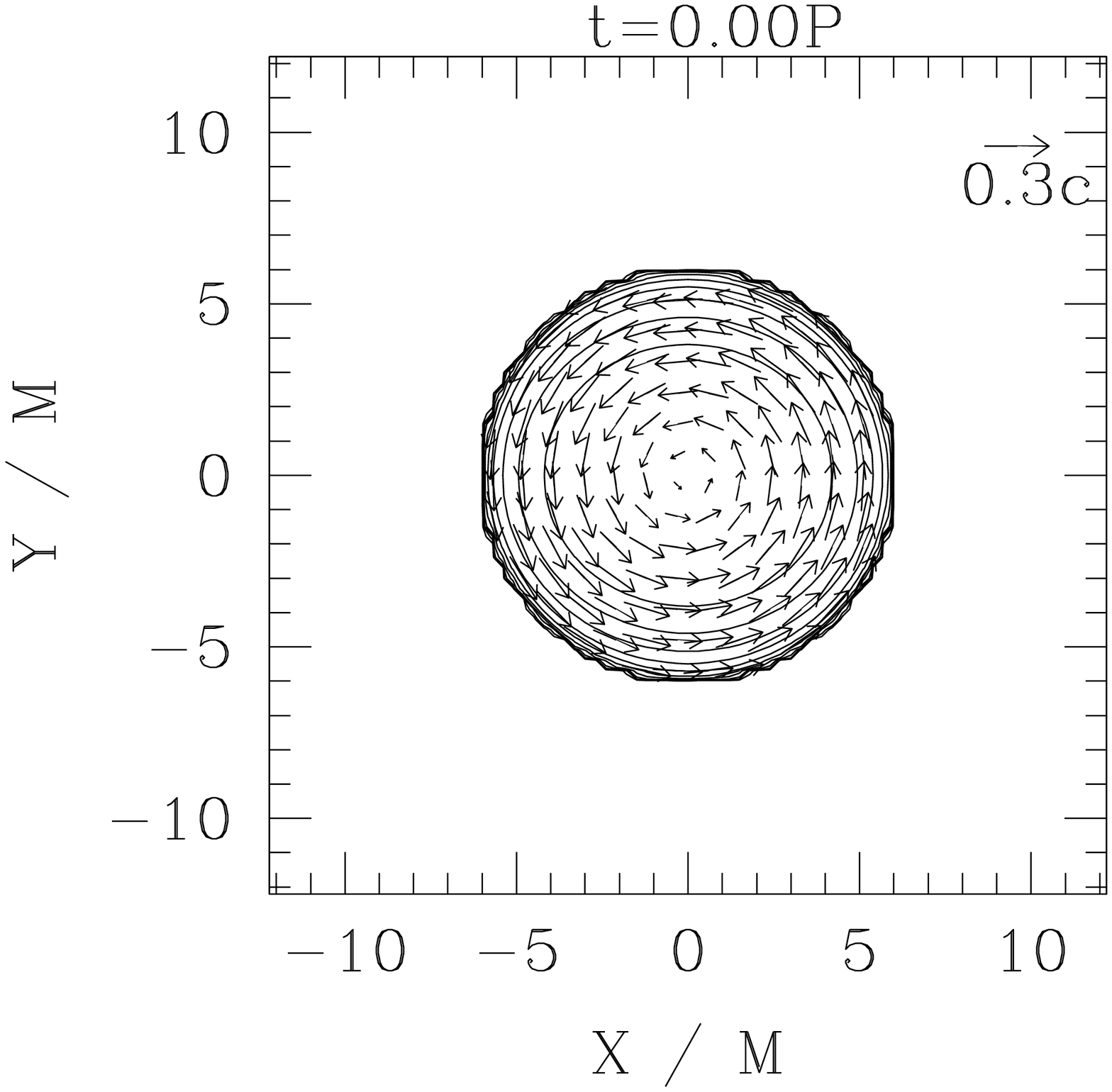}
\epsfxsize=2.5in
\leavevmode
\epsffile{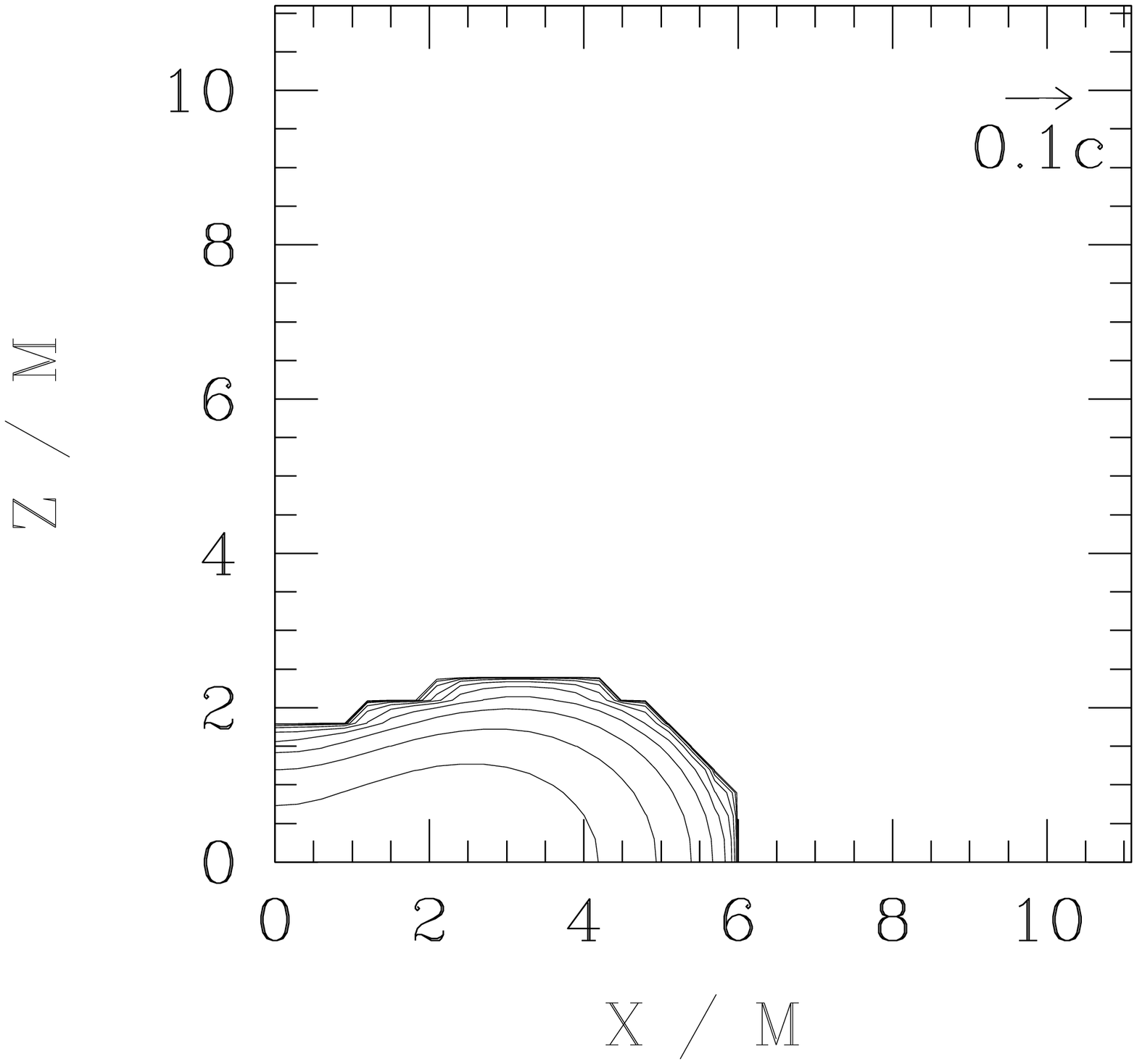}\\
\epsfxsize=2.5in
\leavevmode
\epsffile{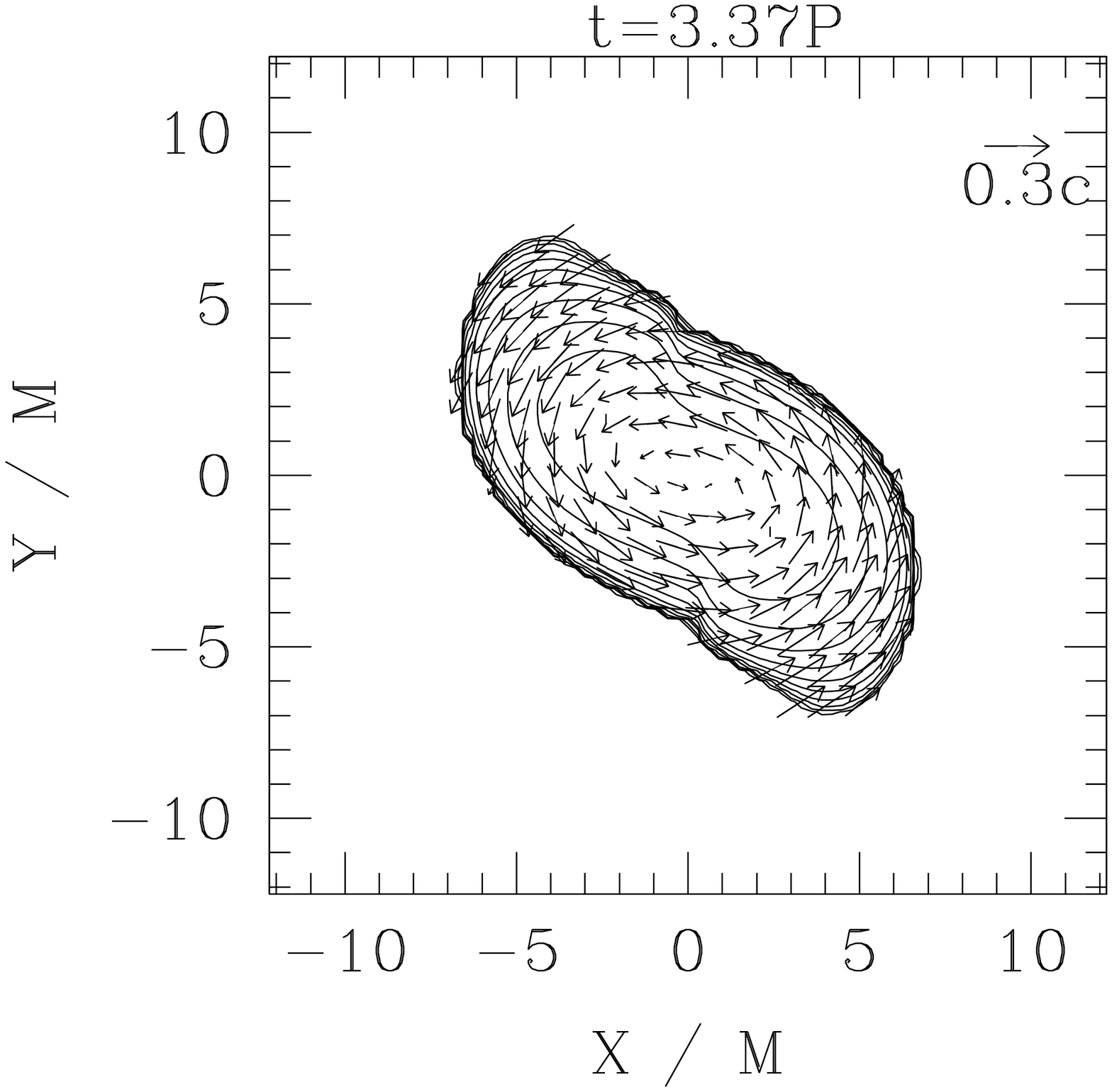}
\epsfxsize=2.5in
\leavevmode
\epsffile{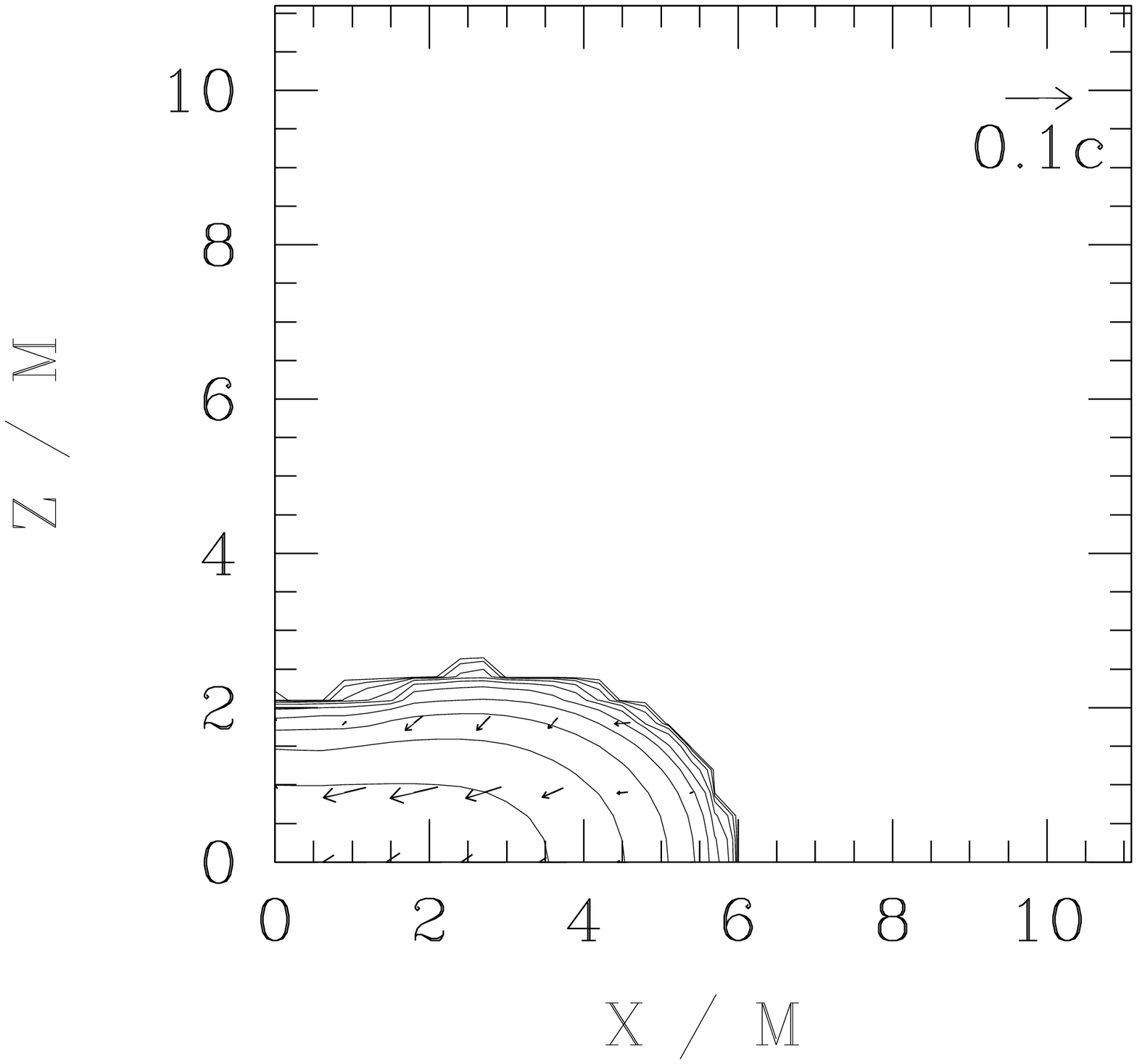}\\
\epsfxsize=2.5in
\leavevmode
\epsffile{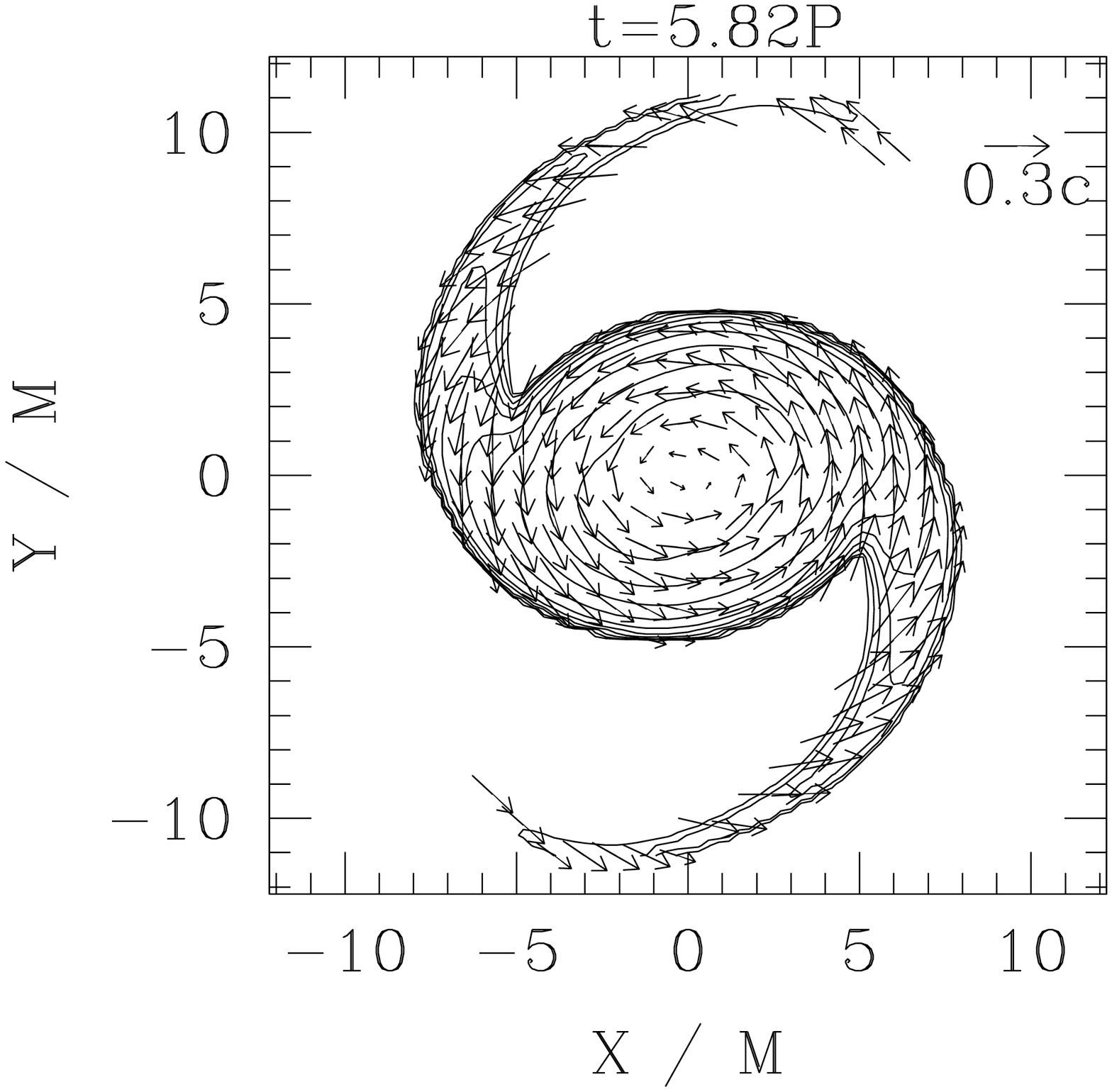}
\epsfxsize=2.5in
\leavevmode
\epsffile{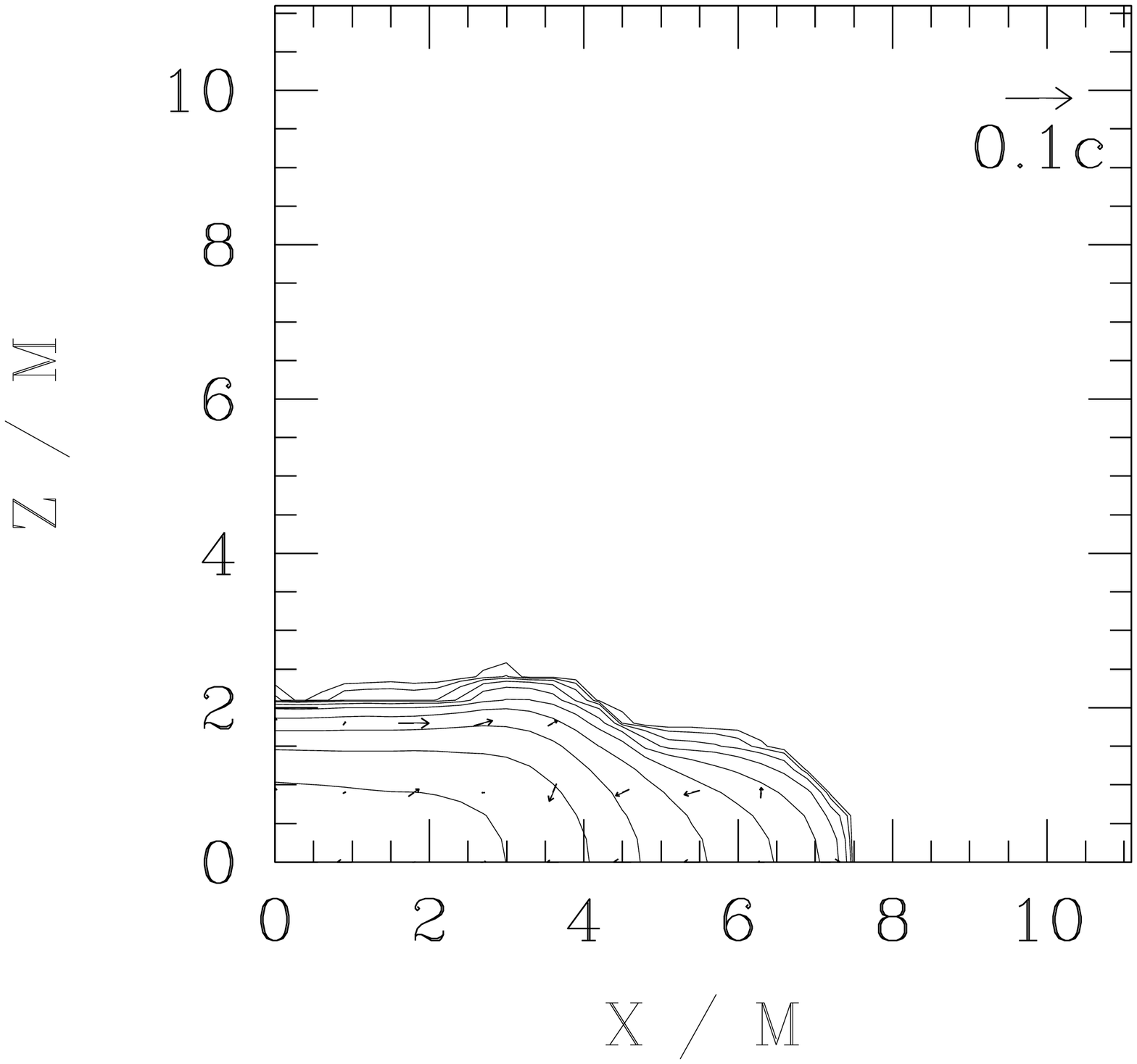}
\end{center}
\caption{Snapshots of density contours for $\rho_*$ and the velocity
flow for $v^i$ in the equatorial plane (left) and in the $y=0$ plane
(right) for the unstable model D2. The contour lines are drawn for
$\rho_*/\rho_{*~{\rm max}}=10^{-0.3j}$ for $j=0,1,2,\cdots,10$ where
$\bar \rho_{*~{\rm max}}$ is 0.126, 0.172 and 0.264 for the three
different times (the corresponding values of $\bar\rho_{\rm max}$ are
0.045, 0.059, and 0.065).  The lengths of arrows are normalized to
$0.3c$ (left) and $0.1c$ (right). The time is shown in units of
$P_{\rm rot}^a$. }
\end{figure}
\clearpage

\begin{figure}[t]
\begin{center}
\epsfxsize=2.5in
\leavevmode
\epsffile{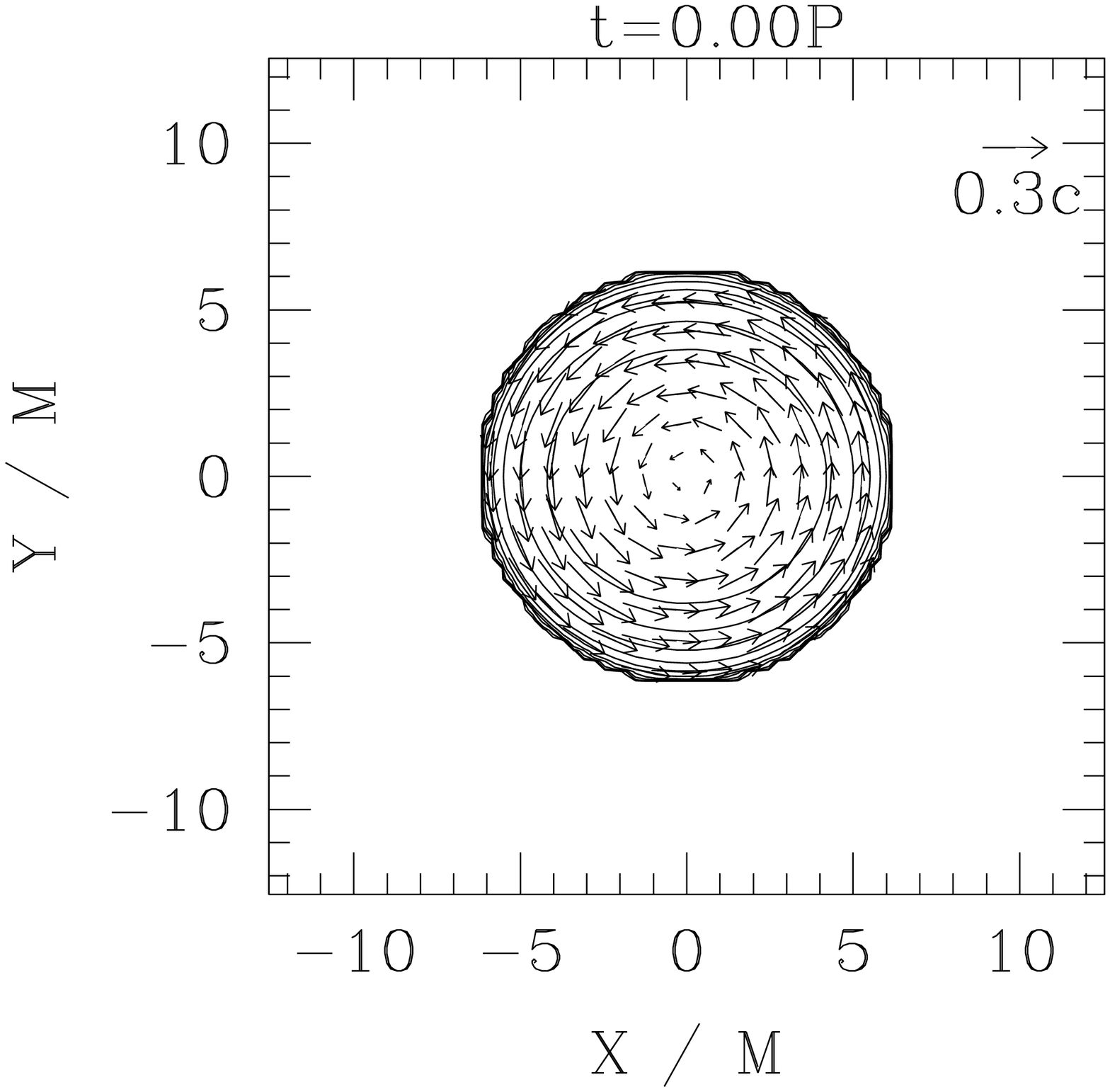}
\epsfxsize=2.5in
\leavevmode
\epsffile{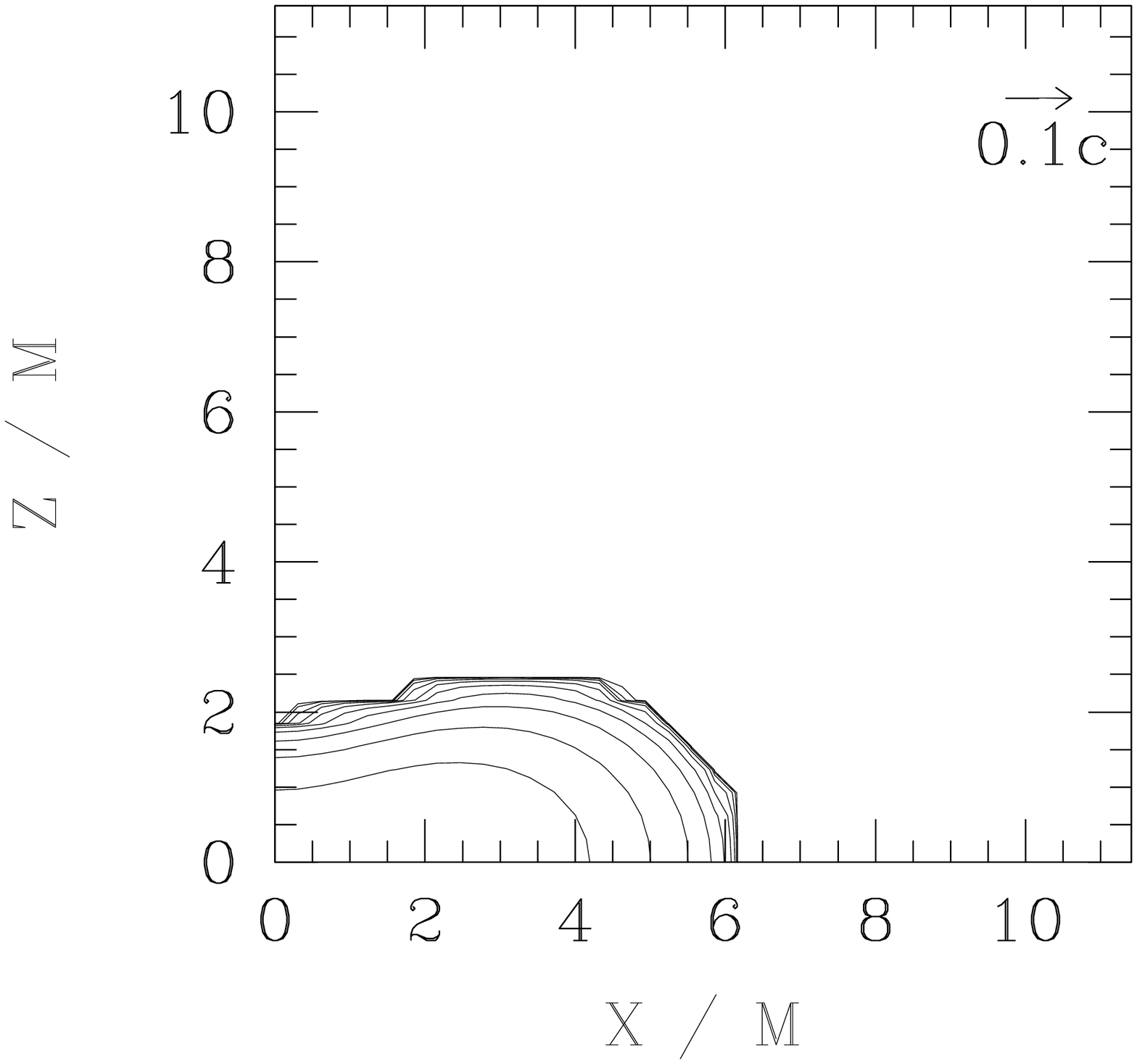}\\
\epsfxsize=2.5in
\leavevmode
\epsffile{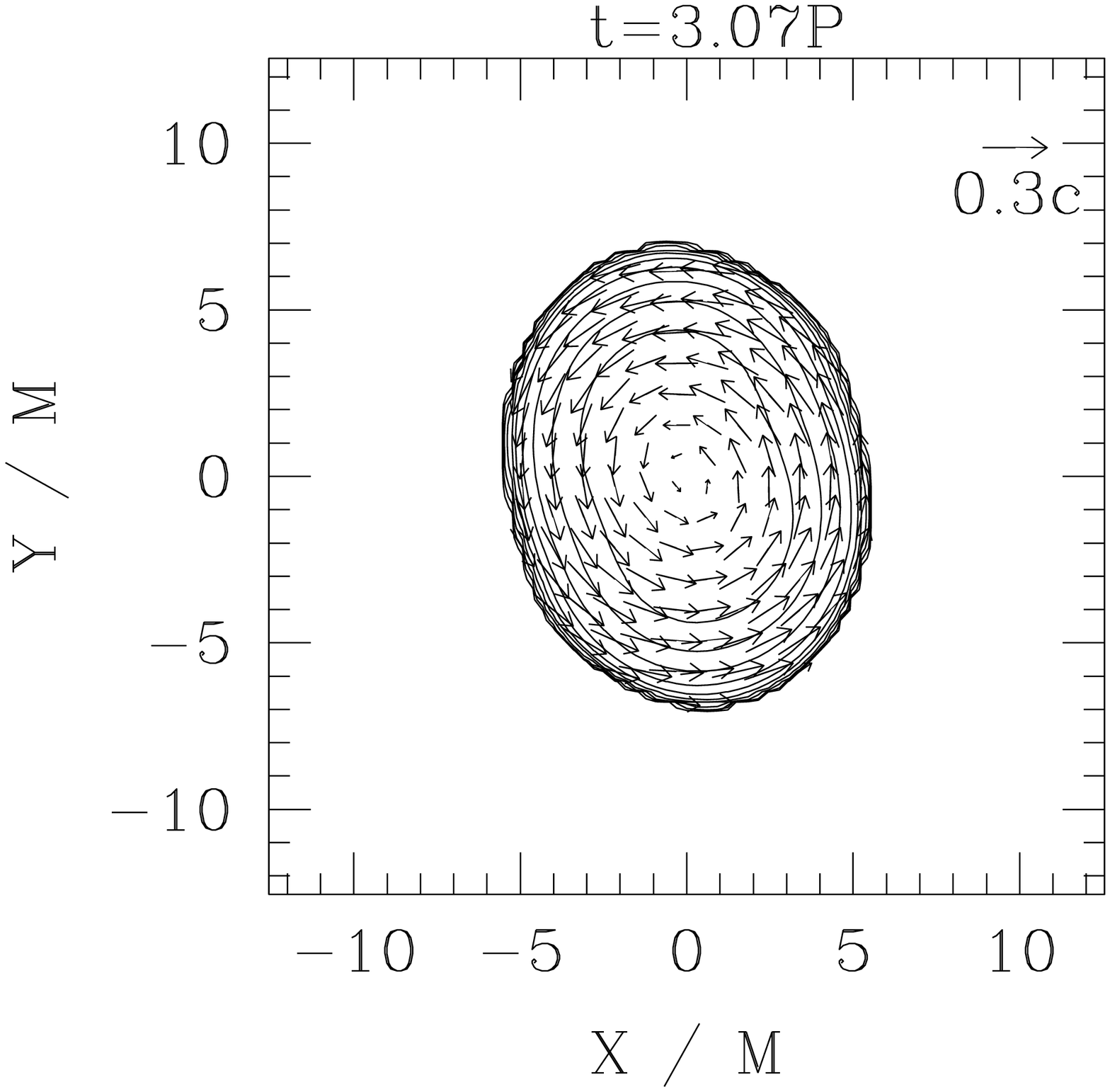}
\epsfxsize=2.5in
\leavevmode
\epsffile{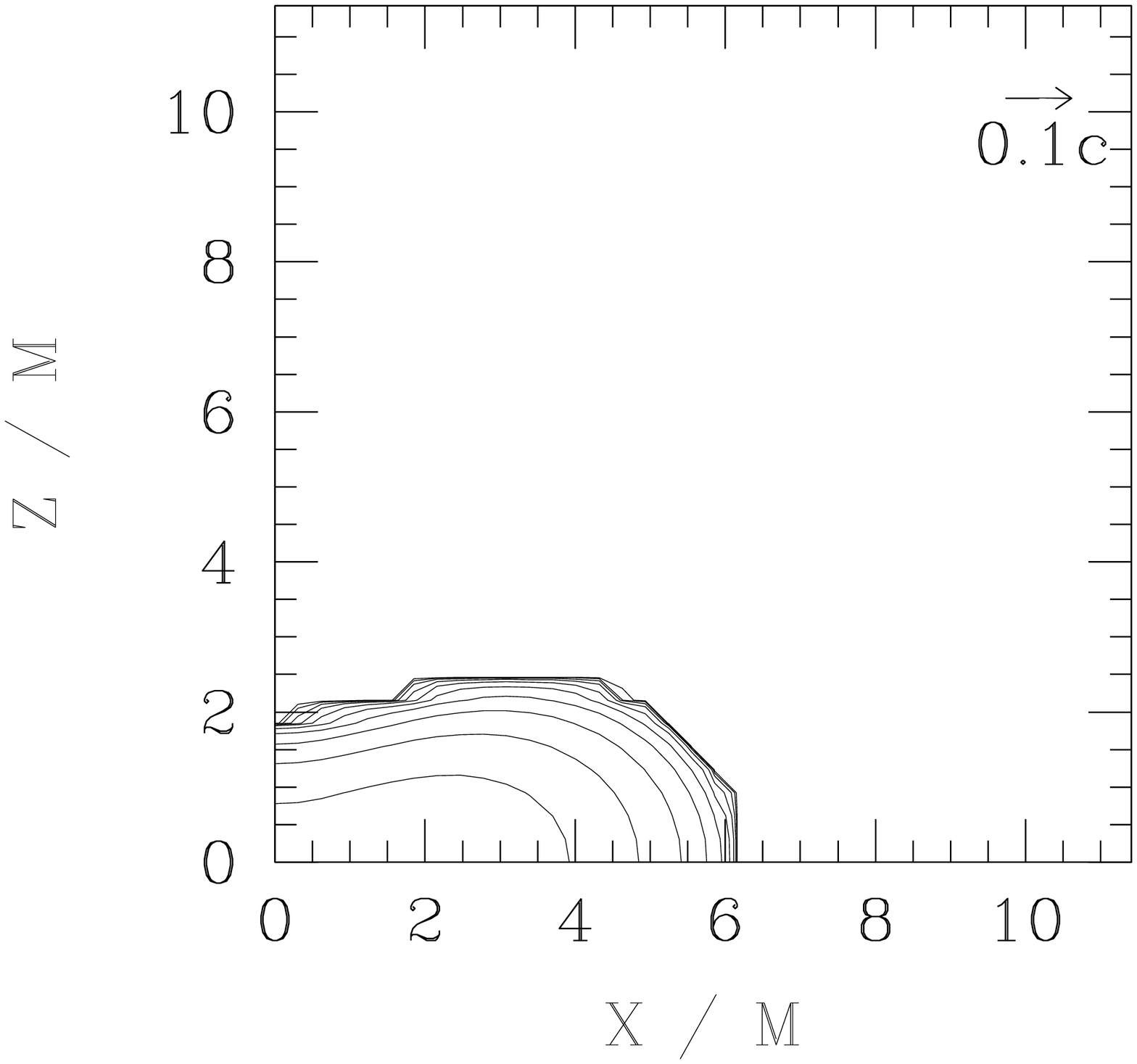}\\
\epsfxsize=2.5in
\leavevmode
\epsffile{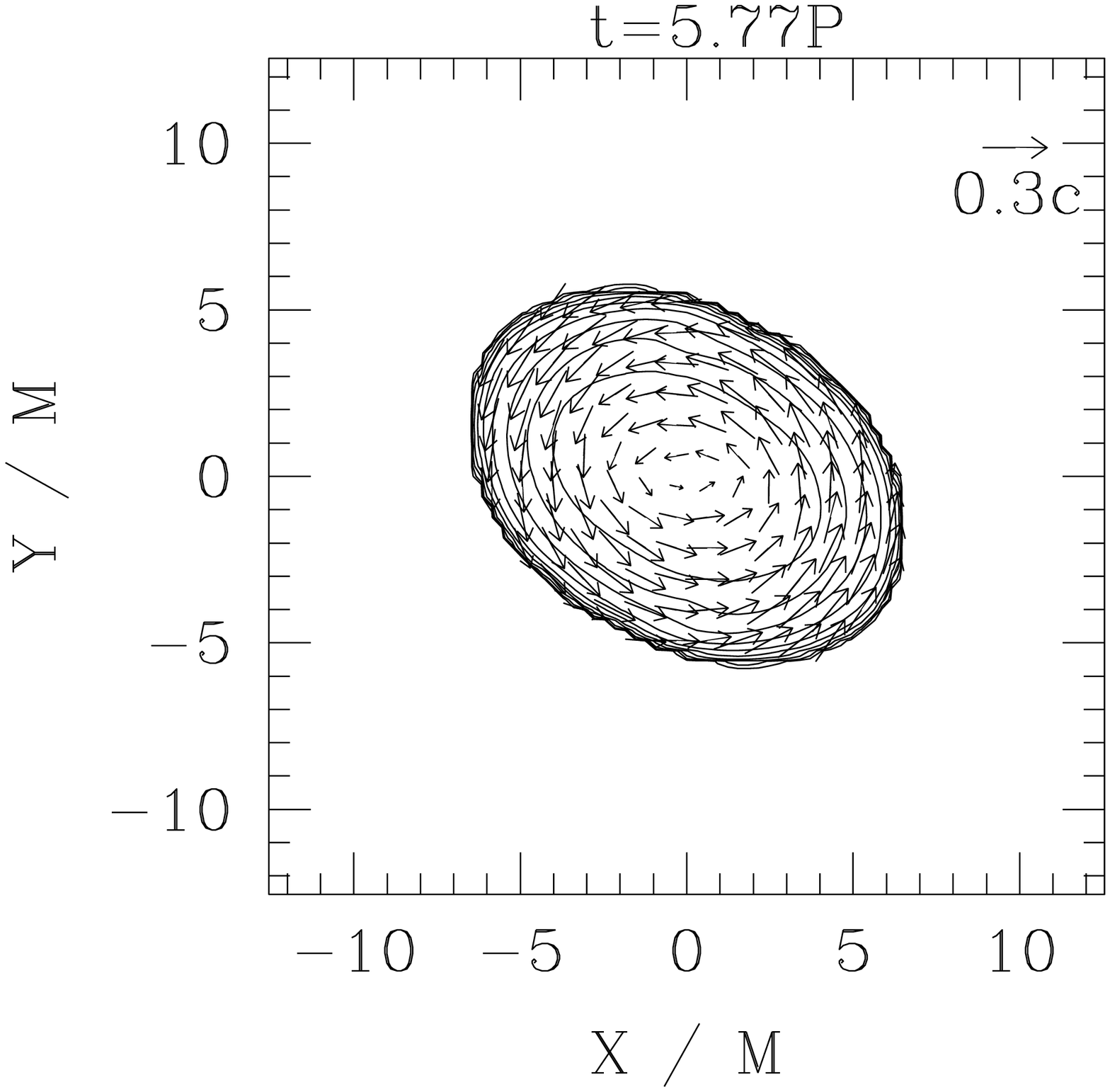}
\epsfxsize=2.5in
\leavevmode
\epsffile{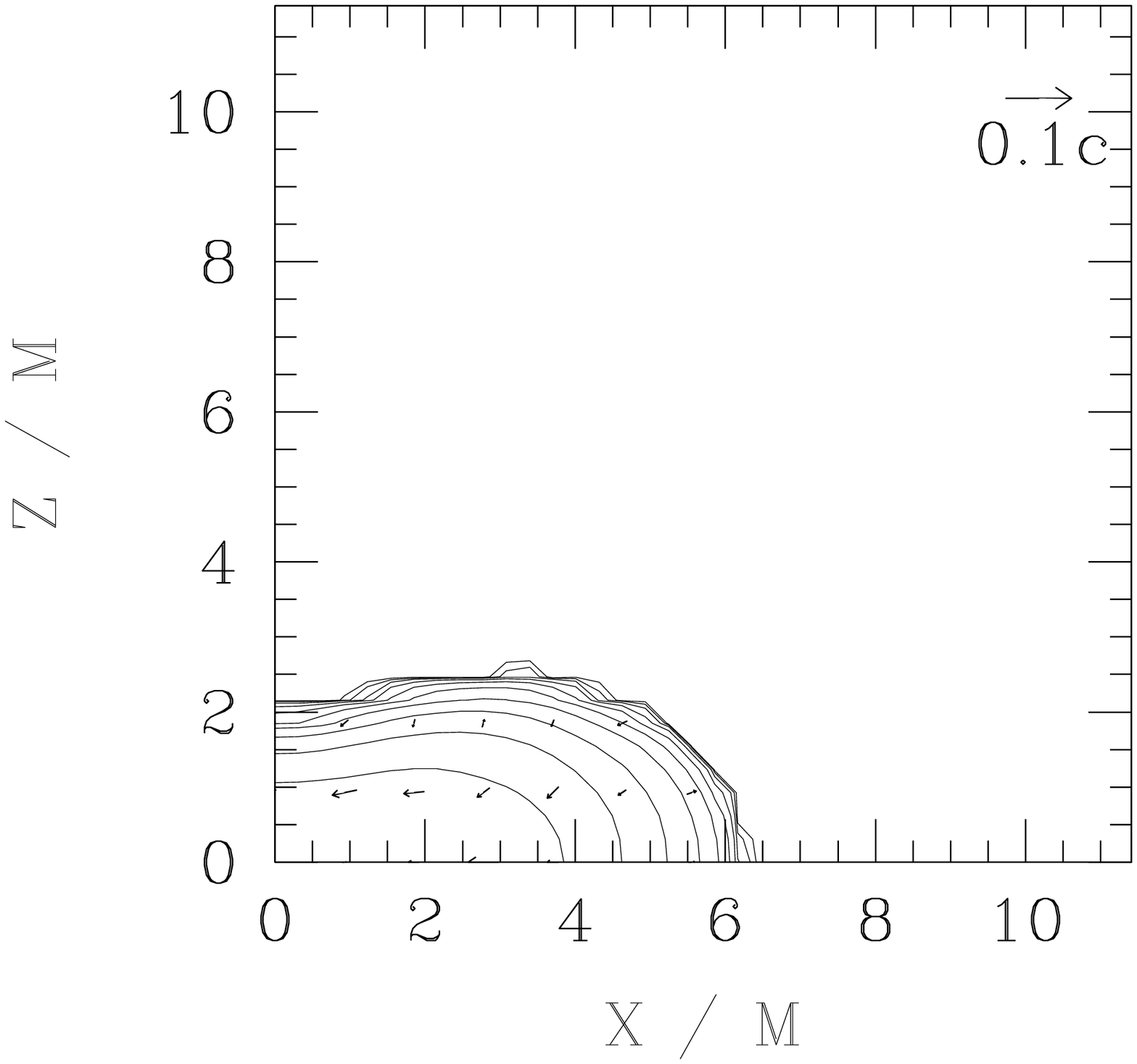}
\end{center}
\caption{Snapshots of density contours for $\rho_*$ and 
the velocity flow for $v^i$ in the equatorial plane (left) and
in the $y=0$ plane (right) for the unstable model D3. The contour lines
are drawn for $\rho_*/\rho_{*~{\rm max}}=10^{-0.3j}$ for $j=0,1,2,\cdots,10$
where $\bar \rho_{*~{\rm max}}$ is 
0.128, 0.152 and 0.176 for the three different
times.  The lengths of arrows are normalized to $0.3c$ (left) and
$0.1c$ (right). The time is shown in units of $P_{\rm rot}^a$. }
\end{figure}

\begin{figure}[t]
\begin{center}
\epsfxsize=4.in
\leavevmode
\epsffile{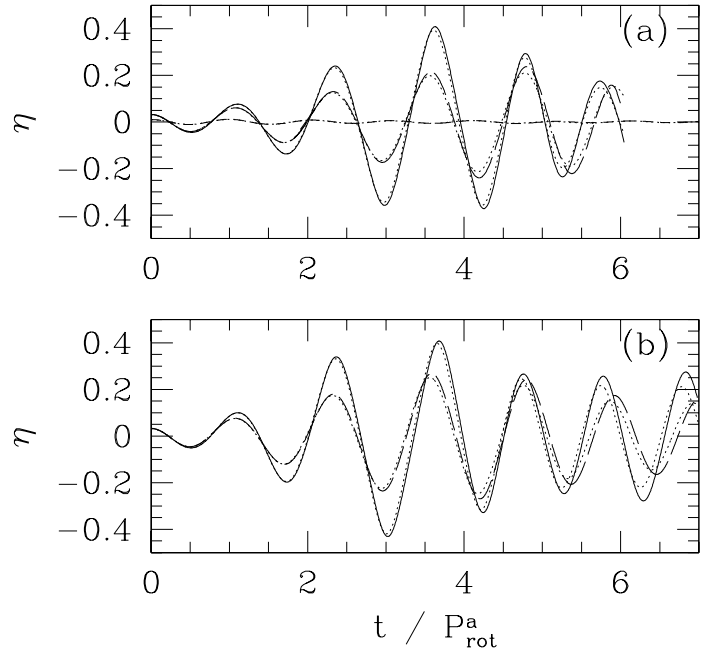}
\end{center}
\caption{The distortion parameter $\eta$ as a function of $t$ (a) for 
models D1 (dashed line), D2 (solid line) and D3 (long-dashed line), and 
(b) for models D6 (solid line) and D7 (long-dashed line). 
The results in a low resolution simulation 
with $101\times 51\times 51$ grid points are shown by the dotted lines. 
}
\end{figure}

\begin{figure}[t]
\begin{center}
\epsfxsize=4.in
\leavevmode
\epsffile{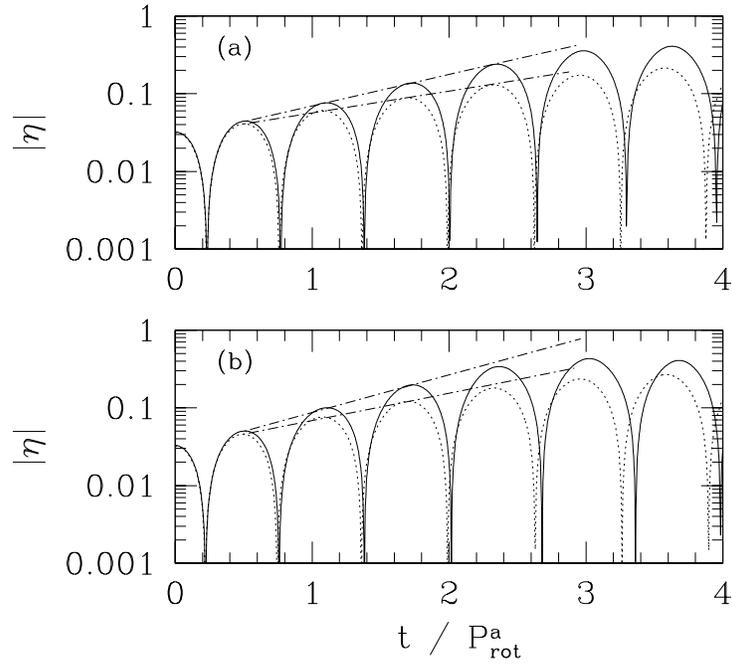}
\end{center}
\caption{$|\eta|$ as a function of $t$ (a) for 
models D2 (solid line) and D3 (dotted line), and 
(b) for models D6 (solid line) and D7 (dotted line). 
The dotted-dashed lines denote the growth time of the 
bar-mode instability. 
}
\end{figure}

\begin{figure}[t]
\begin{center}
\epsfxsize=4.in
\leavevmode
\epsffile{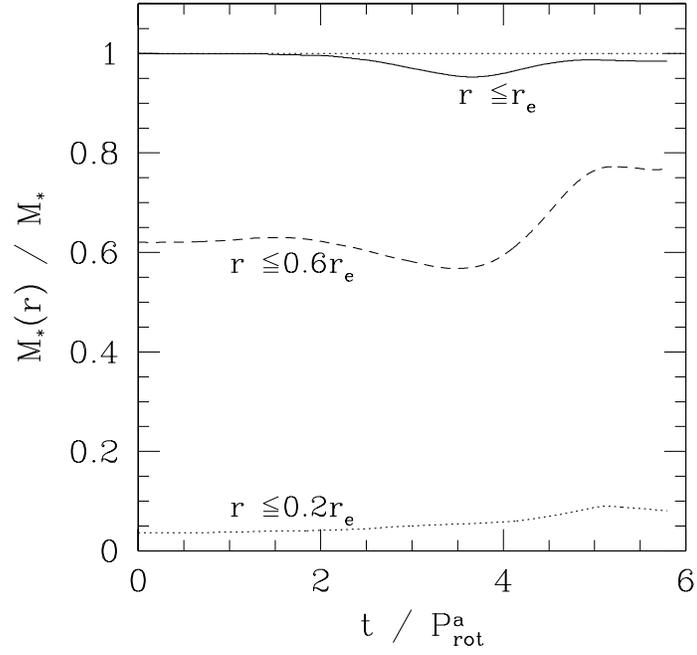}
\end{center}
\caption{The fraction of the rest mass inside a coordinate 
radius as a function of $t$ for the unstable model D2. }
\end{figure}

\begin{figure}[t]
\begin{center}
\epsfxsize=4.in
\leavevmode
\epsffile{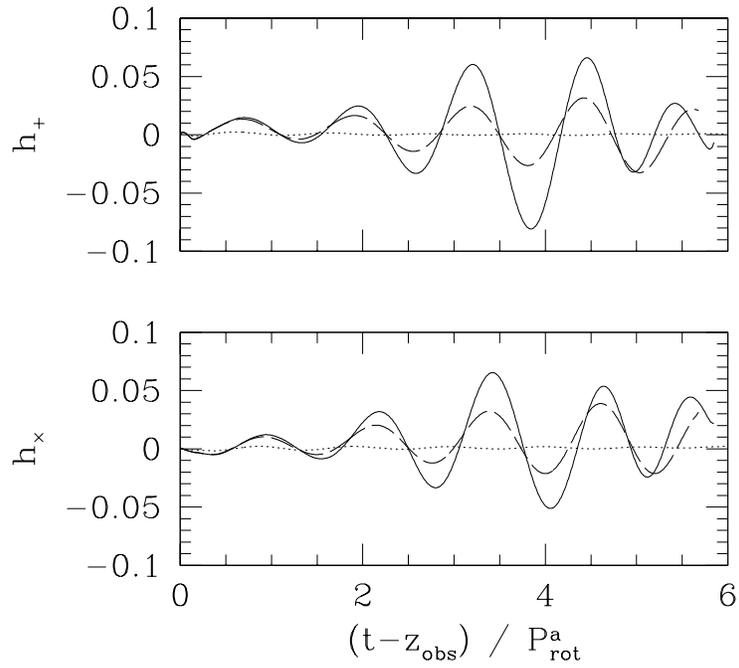}
\end{center}
\caption{$h_+$ and $h_{\times}$ 
as a function of a retarded time 
$t-z_{\rm obs}$ for 
stars  D1 (dotted line), D2 (solid line) and D3 (dashed line).}
\end{figure}

\end{document}